\documentstyle[12pt]{article}
\def\bbox{{\,\lower0.9pt\vbox{\hrule \hbox{\vrule height 0.2 cm

\hskip 0.2 cm

\vrule  height 0.2 cm}\hrule}\,}}

\def\bbox{{\,\lower0.9pt\vbox{\hrule \hbox{\vrule height 0.2 cm

\hskip 0.2 cm

\vrule  height 0.2 cm}\hrule}\,}}
\begin{document}
\setlength{\unitlength}{1mm}
\title{{\hfill {\small } } \vspace*{2cm} \\
Statistical Mechanics, Gravity, and Euclidean Theory}
\author{\\
Dmitri V. Fursaev \date{}}
\maketitle
\noindent  {
{\em Joint Institute for
Nuclear Research, Bogoliubov Laboratory of Theoretical
Physics, \\
141 980 Dubna, Russia}
\\
\\
e-mail: fursaev@thsun1.jinr.ru
}
\bigskip

\begin{abstract}
A review of computations of free energy for Gibbs states on stationary
but not static gravitational and gauge backgrounds is given.  On these
backgrounds
wave equations for free fields are reduced to eigen-value problems
which depend non-linearly on the spectral parameter.  We present a
method to deal with such problems.  In particular, we demonstrate
how some results of the spectral theory of second order elliptic
operators, such as heat kernel asymptotics, can be extended to a class
of non-linear spectral problems.  The method is used to trace down the
relation between the canonical definition of the free energy based on
summation over the modes and the covariant definition given in
Euclidean quantum gravity.  As an application, high-temperature
asymptotics of the free energy and of the thermal part of the
stress-energy tensor in the presence of rotation are derived. We
also discuss statistical mechanics in the presence of Killing horizons
where canonical and Euclidean theories are related in a non-trivial
way.
\end{abstract}

\bigskip
\bigskip
\bigskip
\bigskip
\bigskip
\bigskip
\noindent
{\it These notes are based on the lectures delivered at
the International Meeting on "Quantum Gravity and Spectral
Geometry" in Naples, Italy (July 2-July 7, 2001).}


\baselineskip=.6cm

\newpage

\noindent
\section{Introduction}
\subsection{Basic facts}
\setcounter{equation}0


Quantum field theory at finite temperatures
appeared some forty years ago \cite{Matsubara}--\cite{KMS}
and now it has many applications
ranging from QCD to physics of the early universe, see
\cite{LW}--\cite{FR} for a review.
These notes concern one particular aspect of this theory,
namely, computation of the effective
action and free energy in external
gravitational and gauge fields.

A finite-temperature field theory is closely related
to Euclidean theory. This fact is now
well known and is exploited in numerous applications
\cite{LW}--\cite{FR}. To begin with
we recall how to see
this relation in a simple way.
Consider a
model of a real scalar field $\phi$ in Minkowski space-time with the
action\footnote{We work
in the system of units where $\hbar=c=1$. Our conventions
for the curvature and metric coincide with the book \cite{MTW:73}.}
\begin{equation}\label{i.10}
I_t[\phi]
=
\frac 12\int_0^t dt'\int d^3x\left[-(\partial_t\phi)^2+(\partial_i
\phi)^2+m^2\phi^2\right]~.
\end{equation}
We study a thermal equilibrium of this field in a finite
volume at non-zero
temperature $\beta^{-1}$, a Gibbs state. The state is determined
by the partition function
\begin{equation}\label{i.1}
Z(\beta)=\mbox{Tr}~ e^{-\beta \hat{H}},
\end{equation}
where $\hat{H}$ is a normally ordered Hamiltonian of the
given model.
Normal ordering guarantees that $Z(\beta)$ is
well-defined.
In general, (\ref{i.1}) also requires
$\hat{H}$ to be an operator bounded
from below (which is true for (\ref{i.10})).
The average $\langle\varphi|
e^{-\beta\hat{H}}|\varphi\rangle$ in some quantum state is
equivalent to a transition amplitude from this state to itself for
the period of time $t=-i\beta$. Such analytical continuation from
the real to pure imaginary time is called the Wick rotation.
One
can define a complete set of states
$|\varphi\rangle$ where the field operator
$\hat{\phi}$ is diagonal and takes the values $\varphi$. Then
matrix elements of the evolution operator  can be represented by a
path integral \cite{FH}
\begin{equation}\label{i.2}
\langle \varphi|e^{-it\hat{H}} |\varphi\rangle={\cal N}_t \int
D\phi ~e^{-iI_t[\phi]}~,
\end{equation}
where ${\cal N}_t$ is a normalization coefficient and
trajectories (in a space of functions) in (\ref{i.2}) begin and
end at $\phi=\varphi$. As a result of the Wick rotation,
$I_{t}[\phi]$ transforms into the functional
\begin{equation}\label{i.11}
I^E_\beta[\phi]=iI_{-i\beta}[\phi]
=
\frac 12\int_0^\beta d\tau\int
d^3x\left[(\partial_\tau\phi)^2+(\partial_i
\phi)^2+m^2\phi^2\right]~.
\end{equation}
The latter is the action for the same model but in a flat
space-time where the metric is positive
definite. One says that such as a space-time has the signature $(++++)$
and calls it Euclidean, as opposed to the physical space-time which
has the signature $(-+++)$ and is
called Lorentzian. Now to get
(\ref{i.1}) from (\ref{i.2}) one has to take the trace over all
possible values $\varphi$. Thus, the partition function can be
written as
\begin{equation}\label{i.3}
\mbox{Tr}~ e^{-\beta\hat{H}}={\cal N}_\beta
\int D\phi
~e^{-I^E_\beta[\phi]}~.
\end{equation}
By construction, the path integral is taken over all possible
closed trajectories going along the Euclidean time $\tau$. This is
equivalent to imposing periodicity on $\tau$ with the period
$\beta$. This result can be easily extended to
non-zero spin fields including fields with Fermi statistics.

The integral in r.h.s. of (\ref{i.3}) is exponentially
suppressed for large values of $\phi$ and this
allows its rigorous definition.
One can also
introduce interaction of the field $\phi$ with an external source.
Then the Euclidean path integral can be used as a generating
functional for Euclidean Green functions
\cite{LW},\cite{Kapusta}. The
Euclidean Green functions
correspond to real-time finite-temperature Green
functions which obey certain boundary conditions known as
Kubo-Martin-Schwinger (KMS) condition \cite{KMS}, see
\cite{LW}--\cite{FR} for more details.

It is instructive to give another derivation of (\ref{i.3}). Let
us introduce the free energy of the system
\begin{equation}\label{i.12}
F(\beta)=-\beta^{-1}\ln Z(\beta).
\end{equation}
To calculate it consider a single-particle
excitation $\phi_\omega(t,x)=e^{-i\omega t}\phi_\omega(x)$ of the
field with a frequency $\omega$. The spectrum of $\omega$ can be
found when $\phi_\omega(t,x)$ is  substituted in the wave-equation
(in Lorentzian space-time). As is easy to see, it yields
an eigen-value problem
\begin{equation}\label{i.16}
H^2\phi_\omega=\omega^2 \phi_\omega,
\end{equation}
\begin{equation}\label{i.16a}
H^2=-\partial_i^2+m^2~.
\end{equation}
The operator
$H=\sqrt{H^2}$ is a relativistic analog of the
quantum-mechanical Hamiltonian. In the considered case of a "system
in a box" the spectrum is discrete and positive.
$F(\beta)$ is well-defined and
can be represented in the form \cite{LaLi}
\begin{equation}\label{i.17}
F(\beta)=\beta^{-1} \sum_{\omega}
\ln\left(1-e^{-\beta\omega}\right),
\end{equation}
which is true for free fields.

Let us also define the Euclidean effective action
\begin{equation}\label{i.13}
W^E(\beta)=-\ln
\int D\phi
~e^{-I^E_\beta[\phi]}=\frac 12 \sum_{\Lambda}\ln
\Lambda~,
\end{equation}
where $\Lambda$ are eigen-values of
the Euclidean operator
$L^E=-\partial_\tau^2-\partial_i^2+m^2$.
The series in (\ref{i.13}) diverges and as usually has to
be regularized
according to some prescription.
The eigen-vectors of $L_E$ are periodic in
$\tau$. They have the form $e^{i\sigma_l\tau}\phi_\omega$ where
$\sigma_l=(2\pi l)/\beta$ and $l$ is an integer.
Hence, $\Lambda=\sigma_l^2+w^2$ where $\omega$ are
defined by (\ref{i.16}).
The quantities
$\sigma_l$ are known in the literature as the Matsubara
frequencies \cite{LW}.

The relation between $F(\beta)$ and $W^E(\beta)$ can be found by
using the identity \cite{GrRy},\cite{F:98}
\begin{equation}\label{i.7}
\ln\left(1-e^{-\beta \omega}\right)= -\frac 12
\lim_{\nu\rightarrow 0}{d \over d\nu}\zeta(\nu|\omega,\beta)
-{\beta \omega \over 2}~,
\end{equation}
\begin{equation}\label{i.8}
\zeta(\nu|\omega,\beta)=\sum_{l=-\infty}^{\infty}
\left[\sigma_l^2+\omega^2\right]^{-\nu}~.
\end{equation}
The $\zeta$-function (\ref{i.8}) is defined at $\nu=0$ by
analytical continuation from the region $Re~\nu>1/2$. By taking
into account (\ref{i.17}) we find that
\begin{equation}\label{i.18}
\beta F(\beta)
=
\lim_{w_0\rightarrow \infty}\sum_{w<w_0}\left[ -\frac 12
\lim_{\nu\rightarrow 0}{d \over d\nu}\zeta(\nu|\omega,\beta)
-{\beta \omega \over 2}\right].
\end{equation}
The first term in r.h.s. of this equation formally coincides with
the determinant of the Euclidean operator, see (\ref{i.13}). The
second term in (\ref{i.18}) is related to the energy of vacuum
fluctuations
\begin{equation}\label{i.19}
E_0=\frac 12 \sum_\omega \omega~.
\end{equation}
Hence we conclude that
\begin{equation}\label{i.20}
\beta F(\beta)=W^E(\beta)-\beta E_0~.
\end{equation}
Although both $W^E$ and $E_0$ diverge their
difference is finite and coincides with the free energy. In what
follows we give more rigorous definition of $W^E$ and $E_0$.
Obviously, (\ref{i.20}) is in agreement with relation (\ref{i.3})
derived by the path-integral method where $\ln {\cal N}_\beta$ is
associated to $\beta E_0$.

In case of constant background fields the Euclidean effective
action is
closely connected to the notion of effective potential
\cite{Jackiw:74} and is a very useful tool in studying phase
transitions in a system.
Relation (\ref{i.20}) can be extended to situations when the
system is placed in an arbitrary
static gravitational field as was
first done in pioneering works by Gibbons \cite{Gibbons:77} and
Dowker and Kennedy \cite{DoKe:78} (see also \cite{Allen} for
discussion of (\ref{i.20})). This opens the possibility to study a
variety of physical phenomena in gravitational fields. In a flat
space-time the vacuum energy $E_0$ depends on the boundary
conditions, typically on the size of the system. In
more general situations $E_0$, like other thermodynamical
characteristics, becomes a functional of external background fields.
Thus, the difference of $W^E$ and $F(\beta)$
is essential.

We discussed a finite-size system. In many
physical situations, however,
one has to consider the thermodynamical
limit when the size becomes infinitely large. Strictly
speaking, in this situation the Gibbs state cannot be realized by
a density matrix in a Fock representation. The partition
function is not defined because the operator $\exp(-\beta H)$ is
not of trace class. As was shown by Haag, Hugenholtz and Winnik
\cite{HHW} the mathematically satisfactory definition of
a Gibbs state in
this case can be based on the algebraic Gel'fand-Naimark-Segal
(GNS) construction. The  state is realized as a
reducible vector state in GNS representation
compatible with KMS
condition.

Although the density matrix cannot be introduced for spatially
infinite systems the free energy (\ref{i.17}) and some other
quantities like Green functions can still be defined. One can
start with  a "system in a box" and consider a suitable limit when
the box is taken away to infinity.
Because of spatial homogeneity
free energy of the considered model
will be proportional to the volume of the box.
Thus,
its  density per unit volume is well-defined. Analogously one can
define the density of $F(\beta)$ when the spatial part of
space-time is a hyperbolic manifold of constant negative
curvature.  This can be done by replacing the summation over the
modes in (\ref{i.17}) with an appropriate integral where the
integration measure is determined by the spectral density of the
operator $H^2$, see \cite{BCVZ:96} for a review.
We will follow a similar method in next sections.

Equation (\ref{i.20}) will be a starting point of our further
discussion. It relates the two definitions of free energy: the
first is canonical, $F^C(\beta)=F(\beta)$. It is based on
(\ref{i.17}) and is called "summation over the modes". It is the
definition which follows from principles of statistical
mechanics and has a clear physical meaning. The second is the
definition of free energy in terms of the Euclidean effective
action
(\ref{i.13}), i.e., as $F^E(\beta)=\beta^{-1}W^E(\beta)$.
The two definitions are related\footnote{In some works $F^E$ is
also defined
without the vacuum energy $E_0$. In this case on static space-times
$F^E$ and $F^C$ coincide.}
as (see (\ref{i.20}))
\begin{equation}\label{main}
F^E(\beta)=F^C(\beta)+E_0.
\end{equation}
Once restricted to static space-times $W^E(\beta)$ and, hence,
$F^E(\beta)$ have a number of attractive properties. i) The classical
action $I^E_\beta$ has a covariant form in Euclidean space-time.
Under a proper choice of integration measure in the functional
integral the effective action
becomes a covariant functional of the background metric,
the property which is especially
important in quantum gravity. ii) Euclidean wave operators $L^E$
are positive elliptic operators. There is a number of rigorous
mathematical results concerning the properties of these operators
\cite{Gilkey:84} which can be used to define $F^E$. iii)
Contribution of the vacuum energy $E_0$ is included in $F^E$. This
is one of the reasons why $F^E$, as distinct from $F^C$, can be
defined as a covariant functional. In principle, like in a
classical theory, by varying $F^E$ over the background metric one
can define the total finite-temperature stress-energy of
a quantum field.

\subsection{Difficulties}
\bigskip

We now discuss situations where the Euclidean definition of
free-energy faces difficulties. Consider the same model as
before and assume that the field is inside a cylinder of radius
$R$. If the cylinder has a finite length we still deal with the
field in a box. Let us define the Gibbs state as a thermal
equilibrium of the field as measured in the frame of reference
which rigidly rotates around the symmetry axes of the cylinder
with the angular velocity $\Omega<R^{-1}$.
The metric in the rotating frame can be obtained from
the Minkowski metric written in  polar coordinates
$(r,\varphi)$ (with the origin at the center of the cylinder)
where the polar angle
$\varphi$ should be replaced
to $\tilde{\varphi}=\varphi-\Omega t$
\begin{equation}\label{1.1}
ds^2=-(1-r^2\Omega^2)dt^2
+
2\Omega r^2dt d\tilde{\varphi}+ r^2d\tilde{\varphi}^2+dr^2+dz^2~.
\end{equation}
The free energy can be represented in the form
(\ref{i.17}) where summation is taken over frequencies
$\tilde{\omega}_l$ of quanta with the angular momentum along the
$z$ axis equal to $l$ ($l$ is an integer). If $\omega_l$ is a
frequency in the non-rotating frame, then
$\tilde{\omega}_l=\omega_l+l\Omega$. This relation is easy to see
by rewriting a single-particle solution with  momentum
$l$ in coordinates (\ref{1.1})
\begin{equation}\label{1.2}
e^{-i\omega_l
t}e^{-il\varphi}\phi_{l,\omega_l}(r,z)=e^{-i\tilde{\omega}_l
t}e^{-il\tilde{\varphi}}\phi_{l,\omega_l}(r,z).
\end{equation}
The partition function of the system is defined by (\ref{i.1})
where $\hat{H}$ is a normally ordered
Hamiltonian operator generating time evolution in
the rotating frame (\ref{1.1}). It is not
difficult to construct a formal Feynman path integral
representation for this partition function and come to equation
(\ref{i.3}). To get the Euclidean action $I^E_\beta$ one has to
start from action $I_t$ in Minkowsky space-time, write $I_t$
in coordinates (\ref{1.1}), go to Euclidean time
$\tau=it$ and impose periodicity on $\tau$ with period
$\beta$. One finds that
\begin{equation}\label{1.3}
I^E_\beta[\phi]=I_{-i\beta}[\phi]
=
\int_0^\beta d\tau \int r^2dr d\tilde{\varphi}
dz[g^{\mu\nu}\partial_\mu \phi
\partial_\nu \phi+m^2 \phi^2]~,
\end{equation}
where $g^{\mu\nu}$ is a contravariant tensor defined as
\begin{equation}\label{1.4}
g^{\mu\nu}p_\mu p_\nu
=
p_\tau^2+2i\Omega p_\tau
p_{\tilde{\varphi}}+(r^{-2}-\Omega^2)p_{\tilde{\varphi}}^2+p_r^2+p_z^2~.
\end{equation}
The important difference as compared to the non-rotating ensemble
is that the form (\ref{1.4}) is complex.
Thus, the Euclidean classical action
in the Feynman path integral (\ref{1.4}) becomes a complex
functional. This results in a difficulty because the corresponding
wave operator in the path integral is not a positive-definite
elliptic operator\footnote{Positive-definite elliptic operator
is an operator whose leading symbol
(\ref{1.4}) is a positive-definite quadratic form for all $p\neq
0$, see \cite{Gilkey:84}.}. The Euclidean
effective action $W^E$ for such complex operators cannot be
rigorously defined and the Feynman path integral (\ref{i.3}) is
nothing else but a formal expression.

This problem can be avoided if we simply make an additional step
and consider analytical continuation to pure imaginary values of
$\Omega=i\breve{\Omega}$. Then the Euclidean action becomes real,
the Feynman integral converges, and the Euclidean
operator $L_E$ is a positive-definite elliptic operator with the
leading symbol
\begin{equation}\label{1.5}
\breve{g}^{\mu\nu}p_\mu p_\nu
=
p_\tau^2-2\breve{\Omega} p_\tau
p_{\tilde{\varphi}}+(r^{-2}+\breve{\Omega}^2)
p_{\tilde{\varphi}}^2+p_r^2+p_z^2~.
\end{equation}
Now we can interpret our theory as truly Euclidean, i.e. on a real
space-time with the signature $(++++)$ and metric
\begin{equation}\label{1.6}
ds_E^2=\breve{g}_{\mu\nu}dx^\mu dx^\nu
=(1+r^2\breve{\Omega}^2)d\tau^2
-2\breve{\Omega} r^2d\tau d\tilde{\varphi}+
r^2d\tilde{\varphi}^2+dr^2+dz^2~.
\end{equation}
By going to the new
coordinate $\varphi=\tilde{\varphi}-\breve{\Omega}\tau$
one can check that (\ref{1.6}) is flat.  Thus, the Euclidean manifold
is a flat $R^4$ space where points $(\tau,\varphi, r,z)$
and $(\tau+\beta,\varphi-\breve{\Omega}\beta, r,z)$
are identified.

This simple flat space example is a good illustration of what
happens when one is trying to relate statistical mechanics on an
arbitrary stationary space-time to the Euclidean theory.
Stationary but not static metrics look as
\begin{equation}\label{1.7}
ds^2=g_{tt}dt^2+2g_{ti}dt dx_i+g_{ij}dx^i dx^j
\end{equation}
where components $g_{\mu\nu}$ do not depend on $t$.
Such metrics appear in different physical
problems. In particular some black hole
or cosmological solutions of the Einstein
equations (e.g., Kerr, Kerr-Newman or G\"odel metrics
\cite{HE})
have form (\ref{1.7}).
The
Wick rotation in this case should be accompanied by an additional
analytical continuation of the parameters of the metric  to get
from (\ref{1.7}) a real Euclidean metric. This procedure implies
complexification of the space-time, the idea which was first suggested
and used by Hartle and Hawking \cite{HH} and Gibbons and Hawking
\cite{GH1},\cite{GH2}. More precisely, this procedure can be
described as follows \cite{Gibbons:79}. One considers a
4-complex-dimensional manifold ${\cal M}_c$ with a complex
contravariant tensor field $g^{\mu\nu}_c$ of type (2,0).
Lorentzian and Euclidean manifolds are defined as real sections of
${\cal M}_c$, i.e., as 2-complex-dimensional submanifolds (slices)
on which the restriction of $g^{\mu\nu}_c$ to the cotangent space
of the slice is a real form with the signature $(-+++)$ or $(++++)$,
respectively. It should be noted that an arbitrary Lorentzian
space-time may not admit a complexification which has a real
section with Euclidean signature and vice versa. We will exclude
these cases from our further discussion.

\bigskip

Although, the described procedure of going to Euclidean theory
makes the Feynman integral well-behaved one loses the original
relation to statistical mechanics.
The problem, thus, is to find out whether there is an analog of
equation (\ref{i.20}) when the space-time is stationary but not
static.

\bigskip

There is another problem which is not present in
static space-times. In static space-times
the spectrum of frequencies $\omega$ is
determined by equation (\ref{i.16}). The latter is obtained from
the wave equation for a field under the substitution
$\phi_\omega(t,x^i)=e^{-i\omega t}\phi_\omega(x)$. Equation
(\ref{i.16}) is the eigen-value problem for $H^2$ which is a
3-dimensional elliptic operator (see (\ref{i.16a})). However, on a
stationary space-time the same substitution results in equation
\begin{equation}\label{1.8}
\left(-g^{tt}(x)\omega^2 +i\omega(-2g^{ti}(x)\partial_i+b(x))
+
g^{ij}(x)\partial_i\partial_j+c(x)\right)\phi_\omega(x)=0,
\end{equation}
where $b(x)$ and $c(x)$ are defined by the metric and parameters
of the model.  This is an eigen-value problem which
depends polinomially on the spectral parameter $\omega$.
Equation (\ref{1.8}) belongs to a class of non-linear spectral
problems which appear in different physical
situations. A typical example of such a problem is
a two-particle scattering
where the scattering potential depends on the energy of particles
\cite{JJ:72}.
The difficulty with (\ref{1.8}) is that its
quantum-mechanical interpretation is
obscure but, what is worse, this equation is not an eigen-value
problem of any operator. Thus, standard results from the theory of
elliptic operators cannot be used here. Certainly, it is possible
to analyze (\ref{1.8}) in each particular case and try to solve it
explicitly. This is not easy,
however\footnote{There is a number of publications in mathematical
literature where spectrum and eigen functions are studied
for one-dimensional non-linear spectral problems similar
in the form to (\ref{1.8}),
see, e.g., \cite{math1},\cite{math2} and
references in \cite{math2}.}.
For instance, in case of
rotating black holes (\ref{1.8}) is rather complicated and the
fact it allows separation of  variables \cite{Carter:68} is a
big luck. For fields with spins 1/2, 1 and 2
the situation gets more complicated, see
\cite{Chandra}.

Similar difficulties occur in quantum
theory in external static gauge fields. In
the presence of electric field the time
component $A_t(x)$ of vector potential is not
zero and time derivatives in wave equations are
covariant derivatives $\partial_t+ie A_t$. To have
after the Wick rotation a positive definite elliptic operator one
has to go to imaginary potential $A_t=iA_\tau$. This requires
complexification of the vector field $A_\mu$. Analogously, like
(\ref{1.8}), the equation for spectrum of frequencies $\omega$ of
single-particle states in this situation is not a standard
eigen-value problem because it contains terms linear in $\omega$.

Finally, we dwell on another difficulty which appears in case of
black holes. A remarkable property of black holes is the Hawking
effect \cite{Hawking:74}: in quantum theory black holes
evaporate by emitting particles with a thermal spectrum. The
corresponding temperature (the Hawking temperature) is
$T_H=\kappa/(2\pi)$ where $\kappa$ is a characteristic of the
gravitational field strength near the black hole horizon called
the surface gravity (for an introduction to black
hole physics see \cite{MTW:73},\cite{FN}). The Hawking effect
implies that a black hole can be in  thermal equilibrium with a
quantum gas if the temperature of the gas coincides with the
Hawking temperature
(for a Schwarzschild
black hole the corresponding state is known as a Hartle--Hawking vacuum
\cite{HH}).

This result can be understood on purely geometrical ground. The
Schwarzschild metric is a static spherically symmetric metric
\begin{equation}\label{1.9}
ds^2=-B(r)dt^2+{dr^2 \over B(r)}+r^2(\sin^2\theta
d\varphi^2+d\theta^2),
\end{equation}
where $B(r)=1-r_h/r$. The horizon is located at $r=r_h$ and is a
2-sphere of radius $r_h=2MG$, where $M$ is the mass of a black
hole, $G$ is the Newton constant.
The Euclidean theory corresponding to a
finite-temperature theory around a Schwarzschild black hole is
defined on a space obtained from (\ref{1.9}) by changing $t$ to
$-i\tau$ where $\tau$ has to be periodic with a period $\beta$. As
is easy to see, near the horizon the Euclidean metric looks as
\begin{equation}\label{1.10}
ds_E^2=\kappa^2\rho^2d\tau^2+d\rho^2+r_h^2(\sin^2\theta
d\varphi^2+d\theta^2),
\end{equation}
where $\rho\simeq 2\sqrt{r_h(r-r_h)}$ and $\kappa=1/(2r_h)$ is the
surface gravity. The coordinates $\rho, \kappa\tau$ in
(\ref{1.10}) behave as polar coordinates on the plane. If
$\beta$ is arbitrary the space with metric
$\kappa^2\rho^2d\tau^2+d\rho^2$ is a cone. The conical singularity
vanishes at $\beta=2\pi/\kappa$. This corresponds exactly to
the Hawking temperature.

If
$\beta\neq\beta_H=T_H^{-1}$ the Euclidean space is not smooth.
Conical singularities are defects of
the geometry where the
curvature has delta-function-like behaviour.
In
the presence of such defects quantum theory acquires new
ultraviolet divergencies \cite{Fursaev:94a}-\cite{CKV:94} in a
form of additional divergent terms in the effective action with a
support on the Euclidean horizon.

The canonical formulation of statistical-mechanics in the presence
of Killing horizons also has non-trivial features. The local
temperature measured at a point with coordinate $r$
is given by the Tolman formula, $T(r)=T_H/\sqrt{B(r)}$.
The factor
$\sqrt{B}$ appears because of the
gravitational blue shift of frequencies near the
horizon with respect to frequencies at infinity.
When $r$ approaches $r_h$ the local temperature becomes
infinitely large. So near a black hole one has a thermal bath at
very large temperature. In this regime masses of fields play no
role and the canonical free energy $F^C$ acquires infrared
divergencies.

As a result, in the presence of horizons the Euclidean and the
canonical free energies look different and finding the relation
between them becomes a problem.

\subsection{Content}
\bigskip

The remaining part of these notes is mainly devoted to two
topics where the results
are new, i.e. to studying non-linear spectral problems
and finding relation between canonical and
Euclidean methods for stationary backgrounds.
In Section 2, we present an approach
which enables one to reduce non-linear spectral problems
to standard
eigen-value problems of some fiducial elliptic operators.
In particular, in section 2.3 we demonstrate that some elements of
the
spectral geometry can be extended to this class of spectral problems.
We introduce an analog of the trace of the heat kernel operator
(a pseudo-trace)
and show that its asymptotic expansion is determined
by local geometrical invariants which generalize
the known heat-kernel coefficients.
We briefly discuss some physical applications such as
high-temperature asymptotics of the free energy and
of the stress-energy tensor including the case of non-zero
rotation.
In Section 3, we use our method
to study relation between statistical mechanics
and Euclidean theory on arbitrary stationary space-times.
Among other advantages, the Euclidean theory
allows for an alternative interpretation
of the results of Section 2.
In Section 4 we discuss possibility to extend
the spectral asymptotics to a larger class of non-linear
spectral problems. In particular, we demonstrate
how it can be done in case of
an external electric field.
The aim of Section 5 is to consider the features which
appear in case of quantum theory in the black hole exterior,
or, more exactly, in case of Killing horizons.

It should be emphasized that we are not trying to present
mathematically rigorous proofs of our statements but rather
use arguments which are closer to physicists.
Thus, some of our conclusions can be considered as conjectures.
We test, however, how they work
in concrete non-trivial examples.
Our discussion will be restricted to case of free scalar
fields. This enables us to avoid unnecessary complications
and concentrate on the main problems. When possible we
give references to results concerning higher spin fields.

\section{Stationary space-times}
\setcounter{equation}0

\subsection{Killing frame}
\bigskip

We begin with some helpful
definitions. Consider a field $\phi$ on a domain $\cal M$
of a $D$-dimensional space-time with a time-like Killing
vector\footnote{By definition the Killing vector is
a solution of equation $\xi_{\mu;\nu}+\xi_{\nu;\mu}=0$.}
field $\xi^\mu$ ($\xi^2<0$). ${\cal M}$ may be a complete
manifold if $\xi^\mu$ is everywhere time-like. In most other cases
$\xi^\mu$ is time-like only in some region. We will study
solutions of field equations in the frame of reference related to
Killing observers whose velocity $u^\mu$ is parallel to $\xi^\mu$
\begin{equation}\label{1a.1}
u^\mu=B^{-1/2}\xi^\mu~,~B=-\xi^2.
\end{equation}
For a Killing observer a solution $\phi_\omega$ carrying the
energy $\omega$ is defined as
\begin{equation}\label{1a.2}
i{\cal L}_\xi \phi_\omega=\omega \phi_\omega~~~
\end{equation}
where ${\cal L}_\xi$ is the Lie derivative along $\xi^\mu$. The
background metric $g_{\mu\nu}$ can be represented as
\begin{equation}\label{1a.6}
g_{\mu\nu}=h_{\mu\nu}-u_\mu u_\nu,
\end{equation}
where $h_{\mu\nu}$ is the projector on the directions orthogonal
to $u_\mu$. Because the
shear and expansion of the family of Killing
trajectories vanish identically the trajectories are characterized
at each point only by their acceleration $w_\mu$ and the rotation
$A_{\mu\nu}$ with respect to a local Lorentz frame \cite{FN}
\begin{equation}\label{1a.3}
w_\mu= u_{\mu;\lambda}u^\lambda~~~,
\end{equation}
\begin{equation}\label{1a.5}
A_{\mu\nu}=\frac 12 h^\lambda_\mu h^\rho_\nu (u_{\lambda;\rho} -
u_{\rho;\lambda}).
\end{equation}
To proceed it is convenient to choose coordinates $x^\mu=(t,x^i)$,
$i=1,D-1$, where $\xi=\partial/\partial t$ and, consequently,
$h_{0\mu}=0$. According with (\ref{1a.1}), (\ref{1a.6}), the
interval on $\cal M$ can be written as
\begin{equation}\label{1a.7}
ds^2=-B(dt+a_idx^i)^2+dl^2~,
\end{equation}
\begin{equation}\label{1a.8}
-{1 \over \sqrt{B}}(u_\mu dx^\mu)=dt+a_idx^i~,
\end{equation}
\begin{equation}\label{1a.8a}
a_i=-{u_i \over \sqrt{B}}~,
\end{equation}
\begin{equation}\label{1a.9}
dl^2=h_{\mu\nu}dx^\mu dx^\nu=h_{ij} dx^i dx^j~.
\end{equation}
The role of
$a_i$ becomes clear under synchronization of clocks
by sending light signals:
the two events with coordinates
$(t,x^i)$ and $(t-a_kdx^k,x^i+dx^i)$ are occuring at the same moment.
The metric $dl^2$ serves to measure the proper distance between
these points.

In coordinates $(t,x^i)$ the only non-zero
components of acceleration (\ref{1a.3}) and rotation (\ref{1a.5})
are
\begin{equation}\label{1a.10}
w_i=\frac 12 (\ln B)_{,i}~~~,~~~ A_{ij}= -\frac 12
\sqrt{B}(a_{i,j}-a_{j,i}).
\end{equation}
In four-dimensional space-time one can define a vector of local
angular velocity
\begin{equation}\label{1a.11}
\Omega_i=\frac 12 \epsilon_{ijk} A^{jk},
\end{equation}
where $\epsilon_{ijk}$ is a totally antisymmetric tensor. The
absolute value of the angular velocity is
\begin{equation}\label{1a.11b}
\Omega=(\Omega_i \Omega^i)^{1/2}=\left(\frac 12 A^{\mu\nu}
A_{\mu\nu}\right)^{1/2}.
\end{equation}

The form of the metric in the Killing frame, equations
(\ref{1a.7}), (\ref{1a.8}), is preserved under arbitrary change of
coordinates $x^i$ provided $h_{ij}$ and $a_i$ transform as a $D-1$
dimensional tensor and a vector. There is also another group of
transformations, which preserves (\ref{1a.7}), (\ref{1a.8}),
namely, $t=t'+f(x)$, $a_i=a_i'-\partial_if(x)$, where $f$ is an
arbitrary function of $x^i$. Under these transformations $a_i$
changes as an Abelian gauge vector field. By considering
single-particle excitations with fixed energy $\omega$
\begin{equation}\label{4.1}
\phi_{\omega}(t,x^i)=e^{-i\omega t}\phi_{\omega}(x^i)~~~,
\end{equation}
one can realize this group of transformations as a local $U(1)$,
$$
\phi_{\omega}(t,x^i)= \phi'_{\omega}(t',x^i)=e^{-i\omega
t'}\phi'_{\omega}(x^i)~~,~~
$$
\begin{equation}\label{6.1}
\phi_{\omega}(x^i)= e^{i\omega f(x)}\phi'_{\omega}(x^i)~~~.
\end{equation}
In this picture, $\omega$ coincides with an "elementary charge".
To quantize in the Killing frame one needs a full set of modes
$\phi_{\omega}(x^i)$. As follows from the above arguments, the
equations which determine $\phi_{\omega}(x^i)$ have a form of $D-1$
dimensional equations for charged fields in external gauge field
$a_i$ on a space with the metric $h_{ij}$. It is important that
covariant properties of the theory in $D$ dimensions guarantee
diffeo- and gauge-covariant form of the $D-1$ dimensional problem.
Such a reduction from $D$ to $D-1$ is analogous to the
Kaluza--Klein procedure which yields the Einstein--Maxwell theory
from higher dimensional gravity. The difference between the two
reductions is that in the standard Kaluza--Klein approach the
"extra" dimensions are compact and the charges are quantized.

\bigskip

Let $\cal B$ be a $D-1$ dimensional
space with metric  $dl^2$, see (\ref{1a.9}).
Consider a point $p$ on $\cal B$ with coordinates
$x^i$ and a vector $V_i$ from the tangent space at $p$.  On $\cal
M$, $p$ corresponds to a trajectory of a Killing observer with the
same coordinates $x^i$.  At any point of the trajectory one can
define a vector $V_\mu$ orthogonal to $u^\mu$ such as $V_i=h_i^\mu
V_\mu$. Suppose that connection $\tilde{\nabla}_i$ on $\cal B$ is
determined by $h_{ij}$. Then the covariant derivative with respect
to this connection can be written as \cite{HE}
\begin{equation}\label{A.4}
\tilde{\nabla}_j V_i= h^{\lambda}_i h^{\rho}_j
V_{\lambda;\rho}~~~,
\end{equation}
where $V_{\mu;\nu}$ is the covariant derivative on $\cal M$ with
respect to the connection defined by $g_{\mu\nu}$. (One can easily
check that $\tilde{\nabla}_k h_{ij}=0$.) Relation (\ref{A.4}) can
be generalized to an arbitrary field on $\cal M$. For instance,
for a scalar field
\begin{equation}\label{4.7}
h_j^\mu\partial_\mu \phi=(\partial_j-a_j \partial_t)\phi \equiv
D_i\phi ~~~,
\end{equation}
for a vector orthogonal to $u^\mu$
\begin{equation}\label{A.42}
h^{\lambda}_i h^{\rho}_j V_{\lambda;\rho}
=(\tilde{\nabla}_j-a_j\partial_t) V_i \equiv D_jV_i~~~,
\end{equation}
where $V_i=h_i^\mu V_\mu$. The time derivative in (\ref{4.7}),
(\ref{A.42}) appears in general because fields on $\cal M$ change
along the Killing trajectory.  If $\phi$ and $V_\mu$ are solutions
with certain frequency, see (\ref{4.1}), then $D_i$ become
covariant derivatives on $\cal B$ in external gauge field $a_i$.
This demonstrates explicitly diffeo- and gauge-covariance of the
theory which are left after the reduction.

\subsection{Kaluza-Klein method}
\bigskip

\noindent
We now present an approach developed by 
Fursaev in \cite{F:01} \footnote{We discuss scalar fields only.
Ref. \cite{F:01} also includes the analysis of the spinor fields
but has a mistake in the definition of a spinor Hamiltonian.
Thus its results for spinor fields require a revision.}.
Consider a real scalar field $\phi$ which satisfies the equation
\begin{equation}\label{4.3}
(-\nabla^\mu\nabla_\mu+V)\phi=0,
\end{equation}
where $V$ is a potential. Suppose that $\cal M$ is
a globally
hyperbolic space-time\footnote{A globally hyperbolic space-time
is a one which admits a Cauchy surface. A Cauchy surface $\Sigma$ is
a space-like hypersurface such that every non-space-like
curve intersects  $\Sigma$ exactly once.
For more details see \cite{HE}.}
and
$\Sigma$ is a Cauchy hypersurface in
$\cal M$. One can define the Klein--Gordon inner product on
$\Sigma$ (see \cite{BiDa:82})
\begin{equation}\label{10.5}
<\phi_1,\phi_2>\equiv\int_{\Sigma} d\Sigma^\mu
j_\mu(\phi_1,\phi_2)~,
\end{equation}
\begin{equation}\label{10.4a}
j_\mu(\phi_1,\phi_2)= -i(\phi^{*}_1 \partial_\mu\phi_2
-\partial_\mu\phi^{*}_1\phi_2).
\end{equation}
If $\phi_1$ and $\phi_2$ are solutions to (\ref{4.3}) the current
$j_\mu$ is divergence free, $\nabla^\mu j_\mu=0$. This property
guarantees that (\ref{10.5}) is independent of the choice of $\Sigma$.

Consider first systems in a finite volume (space $\cal B$
is compact).
Then the problem is
formulated as follows: one has to find a complete set of
solutions
$\phi_\omega(t,x^i)$ to (\ref{4.3}) with certain frequencies
$\omega$ which are ortho-normalized with respect to the product
(\ref{10.5}). Frequencies $\omega$ are energies of
single-particle excitations. The spectrum of $\omega$
can be used to define the
canonical free energy $F^C$ in the Gibbs state by (\ref{i.17}).

Let us demonstrate how finding the spectrum can be reduced to a
standard eigen-value problem for an elliptic operator. By using
the results of the preceding section the wave operator in the
Killing frame (\ref{1a.6}) can be represented as
\begin{equation}\label{4.8}
\nabla^\mu\nabla_\mu=\tilde{g}^{\mu\nu}D_\mu D_\nu,
\end{equation}
\begin{equation}\label{4.9}
D_\mu=\tilde{\nabla}_\mu-a_\mu \partial_t,
\end{equation}
where $a_\mu dx^\mu=a_i dx^i$. The connections
$\tilde{\nabla}_\mu$ are determined on some space $\tilde{\cal M}$
with metric
\begin{equation}\label{4.10}
d\tilde{s}^2=\tilde{g}_{\mu\nu}dx^\mu dx^\nu=-Bdt^2+dl^2.
\end{equation}
Relation between $\tilde{\cal M}$ and $\cal M$ becomes transparent
when comparing (\ref{4.10}) with (\ref{1a.7}). We will call
$\tilde{\cal M}$ and $a_\mu$ the fiducial space-time and the
fiducial gauge potential, respectively. Let us consider now a
scalar field $\phi^{(\lambda)}$ on $\tilde{\cal M}$ which obeys
the equation
\begin{equation}\label{4.11}
(-\tilde{g}^{\mu\nu} (\tilde{\nabla}_\mu+i\lambda a_\mu)
(\tilde{\nabla}_\nu+i\lambda a_\nu)+V)\phi^{(\lambda)}=0,
\end{equation}
where $\lambda$ is a real parameter. Because (\ref{4.11}) does not
contain terms linear in the time derivatives, it takes the
following form
\begin{equation}\label{4.12}
H^2(\lambda)\phi^{(\lambda)}_{\omega}(x^i)=\omega^2(\lambda)
\phi_{\omega}^{(\lambda)}(x^i)
\end{equation}
on functions $\phi^{(\lambda)}_\omega(t,x^i) =e^{-i\omega
t}\phi^{(\lambda)}_\omega(x^i)$. The operator $H(\lambda)$ has the
meaning of a relativistic Hamiltonian for the field
$\phi^{(\lambda)}_\omega(x^i)$ on $\cal B$. It takes the
simplest form after the transformation (which does not change the
spectrum)
\begin{equation}\label{2.8}
\bar{H}(\lambda)=e^{-{D-2 \over 2}\sigma} H(\lambda) e^{{D-2 \over
2}\sigma},
\end{equation}
\begin{equation}\label{2.11}
\bar{H}^2(\lambda)=-\bar{h}^{ij}(\bar{\nabla}_i+i\lambda a_i)
(\bar{\nabla}_j+i\lambda a_j)+\bar{V},
\end{equation}
\begin{equation}\label{2.10}
e^{-2\sigma}=-\xi^2=B.
\end{equation}
Connections $\bar{\nabla}_i$ correspond to fields on a $D-1$ 
dimensional space
$\bar{\cal B}$ conformally related to $\cal B$
\begin{equation}\label{2.12}
d\bar{l}^2=\bar{h}_{ij}dx^i dx^j=e^{2\sigma} dl^2.
\end{equation}
The "potential term" in (\ref{2.11}) is
\begin{equation}\label{2.13}
\bar{V} =B\left[V+{D-2 \over 2}(\nabla^\mu w_\mu -{D-2 \over 2}
w^\mu w_\mu)\right],
\end{equation}
where $w_\mu$ is the acceleration (\ref{1a.3}). One can arrive at
(\ref{2.11})  by doing
a conformal transformation in equations
(\ref{4.3}), (\ref{4.11}).
Under this transformation the physical metric
$g_{\mu\nu}$ changes to $\bar{g}_{\mu\nu}=g_{\mu\nu}/B$ and the
Killing vector on the rescaled space has the unit norm,
$\xi^2=-1$.

By taking into account (\ref{2.11}) we conclude that (\ref{4.12})
is the standard eigen-value problem in a finite volume for a
positive-definite elliptic operator. The spectrum of
$H^2(\lambda)$ is discrete and bounded from below
\cite{Gilkey:84}. For simplicity we assume that the spectrum is
strictly positive.

Because the fiducial spectrum depends on $\lambda$ we can find the
physical spectrum from the constraint
\begin{equation}\label{constr}
\chi(\lambda,\omega)=0,~\lambda>0,
\end{equation}
\begin{equation}\label{constra}
\chi(\lambda,\omega)=-\omega^2(\lambda)+\lambda^2.
\end{equation}
If $\lambda$ obeys (\ref{constr}) we denote the corresponding
solution to (\ref{4.11}) as $\phi^{(\omega)}_\omega(t,x^i)
=e^{-i\omega t}\phi^{(\omega)}_\omega(x^i)$. As follows from
(\ref{4.8})--(\ref{4.12}), (\ref{constr}) this field is
also a solution
to physical equation (\ref{4.3}), i.e. one can write
\begin{equation}\label{constr1}
\phi_\omega(t,x^i)=C_\omega\phi^{(\omega)}_\omega(t,x^i),
\end{equation}
where $C_\omega$ is a normalization coefficient. Therefore, the
relativistic eigen-value problem can be reduced to the standard
eigen-value problem for a one-parameter family of elliptic
operators $H^2(\lambda)$ and to solving (\ref{constr}).
We call this method of finding solutions to (\ref{4.3})
the Kaluza--Klein (KK) method.

To fix $C_\omega$
we have to analyze the inner products. The
fiducial fields obey equations (\ref{4.11})  which dictate a
different form of the corresponding vector current and the product
\begin{equation}\label{10.6a}
\tilde{j}_\mu(\phi_1,\phi_2)=
-i(\phi^{*}_1 (\partial_\mu+i\lambda a_\mu)\phi_2
-(\partial_\mu-i\lambda a_\mu)\phi^{*}_1\phi_2),
\end{equation}
\begin{equation}\label{10.7}
(\phi_1,\phi_2)\equiv\int_{\tilde{\Sigma}} d\tilde{\Sigma}^\mu
\tilde{j}_\mu(\phi_1,\phi_2).
\end{equation}
The connection between (\ref{10.5}) and (\ref{10.7})
can be established as follows. (Without loss
of generality we put $B=-\xi^2=1$ because one can always reduce
the problem to this form by conformal rescaling.) Choose $\Sigma$ and
$\tilde{\Sigma}$ 
as constant time hyper-surfaces\footnote{For fiducial space-time
$\tilde{\cal M}$ metric induced on $\tilde{\Sigma}$
coincides with metric on $\cal B$.}. Then
\begin{equation}\label{10.8}
<\phi_\omega,\phi_\sigma>=\int_{\Sigma}\sqrt{h}d^{D-1}x
\left[(\omega+\sigma) \phi^{*}_\omega \phi_{\sigma}+
 i\phi^{*}_\omega a^i(\nabla_i +i\sigma
a_i)\phi_\sigma-i(\nabla_i-i\omega a_i)\phi^{*}_\omega
a^i\phi_\sigma \right],
\end{equation}
\begin{equation}\label{10.9}
(\phi^{(\lambda)}_\omega,\phi^{(\lambda)}_\sigma)=(\omega+\sigma)
\int_{\tilde{\Sigma}}\sqrt{h}d^{D-1}x
(\phi^{(\lambda)}_\omega)^{*}\phi^{(\lambda)}_{\sigma},
\end{equation}
where $h=\det h_{ij}$, and $h_{ij}$ is the metric on
$\cal{B}$. Indices $i,j$ are raised with the help of $h^{ij}$.
Note that up to a multiplier (\ref{10.9}) coincides with standard product
on space of square integrable functions $L^2$. Equations
(\ref{10.8}), (\ref{10.9}) can be used to write
$$
<\phi_\omega,\phi_\sigma>=(\phi_\omega,\phi_\sigma)
$$
\begin{equation}\label{B.2}
+\int_{\Sigma} \sqrt{h}d^{D-1}\left(i\phi_\omega^{*}a^i
(\nabla_i+i\sigma a_i)\phi_\sigma
 -ia^i(\nabla_i-i\omega a_i)
\phi_\omega^{*}\phi_\sigma\right).
\end{equation}
When $\omega\neq\sigma$
one can use the identity
\begin{equation}\label{B.3}
H^2(\omega)-H^2(\sigma)
=(\omega-\sigma)(-2ia^i\nabla_i- i\nabla_i a^i+
(\omega+\sigma)a^ia_i).
\end{equation}
to rewrite (\ref{B.2}) as
$$
<\phi_\omega,\phi_\sigma>=(\phi_\omega,\phi_\sigma)
$$
\begin{equation}\label{B.4}
+{1 \over 2(\sigma^2-\omega^2)}\left[ (\phi_\sigma,
(H^2(\omega)-H^2(\sigma))\phi_\omega)^{*} 
 +(\phi_\omega, (H^2(\omega)-H^2(\sigma))\phi_\sigma)\right].
\end{equation}
The r.h.s. of (\ref{B.4}) vanishes if $H^2(\lambda)$ is Hermitean,
which we assume to be the case.
(\ref{B.4}) demonstrates that
physical modes $\phi_\omega$, $\phi_\sigma$ obtained from fiducial modes
under restriction (\ref{constr}) are
orthogonal with respect to (\ref{10.8}) when $\omega\neq\sigma$.

Consider now the modes with equal frequencies, $\phi_\omega$ and
$\tilde{\phi}_\omega$. (\ref{B.2}) can be written as
\begin{equation}\label{B.5}
<\tilde{\phi}_\omega,\phi_\omega>=(\tilde{\phi}_\omega,\phi_\omega)
+i\int_{\Sigma} \sqrt{h}d^{D-1}x
\tilde{\phi}_\omega^{*}\left[2a^i(\partial_i+i\omega a_i)
+\nabla^i a_i\right]\phi_\omega.
\end{equation}
By differentiating both sides of (\ref{4.12})
with respect to $\lambda$
one finds
\begin{equation}\label{C.2}
\partial_\lambda
H^2(\lambda)\phi^{(\lambda)}_\omega+H^2(\lambda)
\partial_\lambda\phi^{(\lambda)}_\omega
=\partial_\lambda\omega^2(\lambda)\phi_\omega^{(\lambda)}
+\omega^2(\lambda)\partial_\lambda\phi_\omega^{(\lambda)},
\end{equation}
\begin{equation}\label{C.3}
\partial_\lambda H^2(\lambda)=-2ia^i(\nabla_i+i\lambda a_i)
-i\nabla^i a_i.
\end{equation}
Relation (\ref{B.5}) can be rewritten by using (\ref{C.2}),
(\ref{C.3}) and definition
(\ref{constr}) as
\begin{equation}\label{C.4}
<\tilde{\phi}_\omega,\phi_\omega>={\chi'(\omega,\omega) \over 2\omega}
(\tilde{\phi}_\omega,\phi_\omega),
\end{equation}
where $\chi'(\omega,\omega)=\partial_\lambda \chi(\lambda,\omega)$
at $\chi(\lambda,\omega)=0$. Because the form
$(\tilde{\phi}_\omega,\phi_\omega)$ is positive-definite it
follows from (\ref{C.4}) that the form determined by the
Klein-Gordon product is positive-definite if
\begin{equation}\label{C.5}
\chi'(\omega,\omega)>0~.
\end{equation}
If $\chi'(\omega,\omega)<0$ for some $\omega$ the corresponding
state has negative Klein-Gordon norm and
should be excluded. We will assume
that condition (\ref{C.5}) is satisfied for any solution of
(\ref{constr}). We will see later that (\ref{C.5}) also plays a
role in the Euclidean formulation of the theory.

If (\ref{C.5}) is satisfied then the orthogonality of
$\tilde{\phi}_\omega$, $\phi_\omega$ with respect to the standard
product implies their orthogonality with respect to
(\ref{C.4}). Finally, if $\phi^{(\omega)}_\omega$ have a unit
norm, then according to (\ref{C.4}),  normalization of
$\phi_\omega$ in (\ref{constr1}) requires
$|C_\omega|^{-2}=\chi'(\omega,\omega) /(2\omega)$. 

To summarize: we have
proved that functions (\ref{constr1}) are a set of
modes orthonormalized
with respect to the Klein--Gordon inner product
provided the fiducial modes are orthonormalized with respect to
(\ref{10.7}).
This fact 
also means that there is no contradiction between orthonormalization
procedures for physical and fiducial modes.
Given this
result, one can go further and study other
propertiers of the set of physical modes including its completeness
\footnote{For one-dimensional
non-linear spectral problems (for Sturm-Liouville operator)
the completeness of eigen-functions is discussed in \cite{math2}.}.

\subsection{Spectral coefficients}
\bigskip

We discuss now properties of the spectrum of single-particle states.
For the class of models considered in section 2.2 one can
define the function
\begin{equation}\label{n4.1}
K(t)=\sum_\omega e^{-t\omega^2},
\end{equation}
where $t>0$. If the space-time is static (\ref{n4.1}) coincides with
the trace of the heat kernel of the operator $H^2$,
\begin{equation}\label{n4.2}
K(t)=\mbox{Tr}~e^{-tH^2}.
\end{equation}
In general, however, $K(t)$ is not a trace of any operator
like (\ref{n4.2}). Let us call it a pseudo-trace.
For an elliptic operator $H^2$ the asymptotic of (\ref{n4.2}) at
small $t$ has the known form. To avoid unnecessary  complications
we assume that the system is defined on a compact
space without boundaries. Then at small $t$
\begin{equation}\label{n4.3}
\mbox{Tr}~e^{-tH^2}\sim {1 \over (4\pi
t)^{(D-1)/2}}\sum_{n=0}^{\infty}a_n t^n.
\end{equation}
The coefficients $a_n$ are called
Hadamard--Minackshisundaram--DeWitt--Seeley coefficients, or
sometimes the heat coefficients. They are local invariant
functionals polynomial in curvatures of the background metric.

The expansion (\ref{n4.3}) plays an
important role in mathematical \cite{Gilkey:84} and
physical applications. It is natural to ask whether there is an analog
of this expansion for pseudo-trace $K(t)$ in general case.

To answer this question let us note that $K(t)$ diverges at
small $t$ as a result of summation over arbitrary
large $\omega^2$. When $t$ is finite there is the exponential
cutoff of the series at $\omega^2\sim 1/t$. Therefore,
asymptotic properties of $K(t)$ are related to the behaviour of
the spectrum at large $\omega^2$, the same property which
determines the asymptotics of the heat kernel in (\ref{n4.3}). For
large $\omega$ the spectrum becomes sufficiently dense and can be
considered continuous. We will use this fact for our purpose.
Define $K(t)$ for operators with continuous  spectrum as
\begin{equation}\label{n4.4}
K(t)=\int_{\mu}^{\infty} \Phi(\omega)d\omega e^{-t\omega^2}.
\end{equation}
Here $\mu$ is the mass gap of the spectrum. To avoid
complications we also assume that
$\mu>0$ and there are no isolated discrete
levels\footnote{These assumptions seem to be not
essential in studying the limit of large $\omega$.}.
The quantity $\Phi(\omega)$ is the spectral density or the total
spectral measure defined on a compact domain $\Sigma_r$ of
a Cauchy hypersurface $\Sigma$ as
\begin{equation}\label{10.3}
\Phi(\omega)= \int_{\Sigma_r} d\Sigma^\mu \sum_l
j_\mu(\phi_{\omega,l},\phi_{\omega,l}),
\end{equation}
where $j_\mu$ is defined in (\ref{10.4a}).
We assume that
$\phi_{\omega,l}$ is a complete set of single-particle solutions
to (\ref{4.3}). Index $l$ enumerates all modes with the same
frequencies. In problems with a continuous spectrum $\Sigma$ has
an infinite volume and normalization of modes is
to be understood in terms of distributions, i.e., as
\begin{equation}\label{10.27}
<\phi_{\omega,k},\phi_{\sigma,l}>=
\delta_{lk}\delta(\omega-\sigma),
\end{equation}
where $\delta(x)$ is the delta-function. The integral on r.h.s. of
(\ref{10.3}) is divergent. This is the divergence of the large
volume or infrared divergence. To avoid the divergence one has to
work with a regularized density obtained by restricting the
integration in (\ref{10.3}) to some compact domain $\Sigma_r$ 
in $\Sigma$ and
then expanding $\Sigma_r$ to $\Sigma$ \footnote{We will
see that this procedure enables one to study local characteristics
of the system
such as the density of the free energy per unit volume or
a stress-energy tensor.}.

The important advantage of the KK-method is that the spectral density
$\Phi(\omega)$ can be related to the spectral density of the
corresponding fiducial problem (\ref{4.12}).  Suppose that
$\phi^{(\lambda)}_{\omega,k}$ are a complete set of
ortho-normalized solutions of (\ref{4.12})
\begin{equation}\label{10.26}
(\phi^{(\lambda)}_{\omega,k},\phi^{(\lambda)}_{\sigma,l})=
\delta_{lk}\delta(\omega-\sigma).
\end{equation}
If the modes are related by
\begin{equation}\label{constr2}
\phi_\omega(t,x^i)=\phi^{(\omega)}_\omega(t,x^i),
\end{equation}
one can show \cite{F:01} by using (\ref{10.8}) and (\ref{10.9})
that the normalization (\ref{10.27}) follows from (\ref{10.26}).

The (regularized) spectral density of fiducial modes is
\begin{equation}\label{10.3a}
\Phi(\omega;\lambda)= \int_{\tilde{\Sigma}_r} d\Sigma^\mu \sum_l
\tilde{j}_\mu(\phi^{(\lambda)}_{\omega,l},\phi^{(\lambda)}_{\omega,l}),
\end{equation}
where $\tilde{j}_\mu$ is defined in (\ref{10.6a})
and $\tilde{\Sigma}_r$ is a compact domain in $\tilde{\Sigma}$.
The definition of $\tilde{\Sigma}_r$ corresponds to
the definition of the domain $\Sigma_r$ in $\Sigma$
(we choose again $\Sigma$ and $\tilde{\Sigma}$ as constant time
hypersurfaces).
Let us introduce also an auxiliary quantity 
\begin{equation}\label{10.28}
\Psi(\omega;\lambda)= \Phi(\omega;\lambda) -{1 \over
4\lambda}\sum_k \left[(\phi^{(\lambda)}_{\omega,k},
\partial_\lambda H^2(\lambda)\phi_{\omega,k}^{(\lambda)})
 +(\phi^{(\lambda)}_{\omega,k},
\partial_\lambda H^2(\lambda)\phi_{\omega,k}^{(\lambda)})^{*}
\right],
\end{equation}
where $\partial_\lambda H^2(\lambda)$ is given in (\ref{C.3}). As
follows from (\ref{B.2}),
\begin{equation}\label{10.29}
\Phi(\omega)= \Psi(\omega;\omega).
\end{equation}
It should be noted that $\Phi(\omega)$ does not coincide with
$\Phi(\omega;\omega)$, as one could naively expect. The
distinction of these two quantities is in different forms of the
inner products. 
Consider now the spectral representation for the heat kernel of
$H^2(\lambda)$
\begin{equation}\label{11.1}
\mbox{Tr}~e^{-tH^2(\lambda)}= \int_{\mu}^\infty
\Phi(\omega;\lambda) e^{-t\omega^2}d\omega,
\end{equation}
where integration in the trace is restricted by $\tilde{\Sigma}_r$.  We
assume that the mass gap of $H^2(\lambda)$ is positive and
does not
depend on $\lambda$ (which is true in a number of physical
problems). Similarly we can define the integral
\begin{equation}\label{10.32}
\int_{\mu}^\infty \Psi(\omega;\lambda) e^{-t\omega^2}d\omega
=\mbox{Tr}\left[\left(1-{1 \over 2\lambda}\partial_\lambda
H^2(\lambda)\right)e^{-tH^2(\lambda)}\right],
\end{equation}
where the r.h.s. is the consequence of (\ref{10.28}). Because the
trace does not depend on the choice of the basis and, hence, on
$\lambda$ one can write (\ref{10.32}) as
\begin{equation}\label{10.33}
\int_{\mu}^\infty \Psi(\omega;\lambda) e^{-t\omega^2}d\omega
=\left(1 + {1 \over 2\lambda t}\partial_\lambda\right)
\mbox{Tr}~e^{-tH^2(\lambda)}.
\end{equation}
The latter relation is equivalent to
\begin{equation}\label{kk12}
\Psi(\omega;\lambda)=\Phi(\omega;\lambda) +{\omega \over
\lambda}\int_{\mu}^{\omega}
\partial_\lambda \Phi(\sigma;\lambda)d\sigma~~~.
\end{equation}
Formula (\ref{kk12}) is our key relation which
together with (\ref{10.29}) enables one to compute the physical
spectral density $\Phi(\omega)$ by using powerful heat kernel
techniques. It is remarkable that (\ref{kk12}) reappears
in the Euclidean theory (see section 3.3).

Consider now a short $t$ asymptotics of (\ref{11.1}) 
\begin{equation}\label{2.4}
\mbox{Tr} e^{-tH^2(\lambda)} \sim {1 \over (4\pi
t)^{(D-1)/2}} \sum_{n=0}^\infty a_n(\lambda) t^n.
\end{equation}
The coefficients $a_n(\lambda)$ are
heat kernel coefficients for the
operator (\ref{2.11}). They are local functionals on 
$\tilde{\Sigma}_r$ and
are even in $\lambda$
(the fiducial theory is $U(1)$ invariant and the heat
coefficients are even functions of charges). The gauge invariance
also guarantees that the coefficients are polynomials in powers of
the Maxwell stress tensor and its derivatives. In our case the
role of the gauge field is played by the vector $a_i$, and
the corresponding Maxwell tensor $F_{ij}=a_{j,i}-a_{j,i}$
is related to the rotation.
In general,
\begin{equation}\label{2.16}
a_n(\lambda)=\sum_{m=0}^{[n/2]}\lambda^{2m}a_{2m,n}~~~,
\end{equation}
where $a_{2m,n}$ do not depend on $\lambda$. The highest power of
$\lambda$ in (\ref{2.16}) can be determined by
dimensional analysis. Coefficients $a_0$ and $a_1$ in (\ref{2.4}) do
not depend on $\lambda$.

The spectral density at high frequencies  can be found from
(\ref{2.4}) by using (\ref{10.33}). One can neglect for simplicity
the mass gap and use inverse Laplace transform in (\ref{10.33}).
It should be noted, however, that for
operators with zero gap one has to take into account the presence
of infrared singularities which come out in (\ref{11.1}) at small
$\omega$. One of the possibilities to avoid this problem is to
use dimensional regularization and formally consider $D$ as a
complex parameter.

It is instructive first to obtain the
asymptotics for the fiducial spectral density at large $\omega$
\begin{equation}\label{2.5}
\Phi(\omega;\lambda)\sim {2 \omega^{D-2} \over
(4\pi)^{(D-1)/2}}\sum_{n=0}^\infty {a_n(\lambda) \over
\Gamma\left({D-1 \over 2}-n\right)} \omega^{-2n}.
\end{equation}
One can easily verify that for complex $D$ substitution of
(\ref{2.5}) in (\ref{11.1}) yields (\ref{2.4}). Another method
how to define (\ref{2.5}) for integer $D$ is discussed in
\cite{Sergio}. For $\Psi(\omega;\lambda)$ relation (\ref{10.33})
results in expansion
\begin{equation}\label{2.5a}
\Psi(\omega;\lambda)\sim {2 \omega^{D-2} \over
(4\pi)^{(D-1)/2}}\sum_{n=0}^\infty {\tilde{a}_n(\lambda) \over
\Gamma\left({D-1 \over 2}-n\right)} \omega^{-2n},
\end{equation}
\begin{equation}\label{2.5b}
\tilde{a}_n(\lambda)=a_n(\lambda)+{1 \over 2\lambda}
\partial_\lambda a_{n+1}(\lambda).
\end{equation}
Finally, by taking into account (\ref{10.29}), (\ref{2.16}),
(\ref{2.5a}), (\ref{2.5b}) one finds
\begin{equation}\label{2.17}
\Phi(\omega)\sim {2 \omega^{D-2} \over
(4\pi)^{(D-1)/2}}\sum_{n=0}^\infty{c_n \over \Gamma\left({D-1
\over 2}-n\right)} \omega^{-2n},
\end{equation}
\begin{equation}\label{2.18}
c_n =\sum_{m=n}^{2n}(-1)^{n-m}{\Gamma\left(m-{D-1 \over 2} \right)
\over \Gamma\left(n-{D-1 \over 2}\right)} a_{2(m-n),m},
\end{equation}
We call $c_n$ the spectral coefficients.
It is remarkable that they are
local functionals expressed in terms of heat-kernel
coefficients of elliptic operators.

Now some comments are in order.
Formula (\ref{2.18})
is valid in any dimension $D$. As for expansions (\ref{2.5})
and (\ref{2.17}), they are finite in even dimension $D=2k$. For
$D=2k+1$ terms in (\ref{2.17}) with $n\geq k$ are formally
zero. This indicates that one should be more careful with infrared
singularities in this case. The proper way of dealing with the
infrared problem is to keep $D$ complex in (\ref{2.17})
till the
last stage of computations. Then for any $D$ some
relevant quantities determined with the help of $\Phi(\omega)$ are
finite except, possibly, a number of standard poles
(see section 2.5).

\bigskip

We return now to the question
about asymptotics of the pseudo-trace $K(t)$. To find
its behavior at small $t$ it is enough to substitute
(\ref{2.17}) in (\ref{n4.4}).
By neglecting the gap we
get the expansion
\begin{equation}\label{n4.5}
K(t)\sim {1 \over (4\pi t)^{(D-1)/2}}\sum_{n=0}^{\infty}c_n t^n.
\end{equation}
As was expected, (\ref{n4.5}) holds both for odd and even $D$
because infrared singularities play no role.
We conjecture that the pseudo-trace expansion
(\ref{n4.5}) has to be equally valid for
continuous and discrete spectra.

\bigskip

The coefficients in leading terms in (\ref{n4.5})
can be immediately computed by using (\ref{2.18}). For instance,
$c_0=a_0$ and
\begin{equation}\label{xx1}
c_1=a_{0,1}+{D-3 \over 2}a_{2,2}.
\end{equation}
Term $a_{2,2}$ is determined by the gauge part of $a_2(\lambda)$,
see (\ref{2.16}),
\begin{equation}\label{11.2}
a_{2,2}=-{1 \over 12}\int_{\Sigma_r}h^{1/2} d^{D-1}x
F^{ij}F_{ij},
\end{equation}
where $F_{ij}=a_{j,i}-a_{i,j}$ is a "Maxwell tensor" of the fiducial
gauge field. Term $a_{0,1}$ coincides with $a_1$ in expansion
(\ref{2.4}) for operator (\ref{2.11}) with $\lambda=0$.

\subsection{Examples}
\bigskip

There are simple examples where our results can be checked.
One example is motivated by studying
a quantum theory in the Einstein universe
$R^1\times S^{D-1}$.
Consider a Gibbs state in this space-time defined as
a thermal equilibrium
in the frame of reference which rigidly
rotates with
angular velocity $\Omega_0$ (with restriction
$\Omega_0<1/\rho$ where $\rho$ is the radius of $S^{D-1}$).
The
spectrum can be easily found for conformal scalar fields
\begin{equation}\label{n5.1}
(-\nabla^2+\xi_D R)\phi=0,
\end{equation}
where $\xi_D=(D-2)/(4(D-1))$ and $R=D(D-1)/\rho^2$ is the scalar
curvature. Suppose for simplicity that $D=3$ and $\rho=1$. The
metric is
\begin{equation}\label{n5.2}
ds^2=-dt^2+\sin^2\theta d\varphi^2+d\theta^2.
\end{equation}
In the rotating frame this metric can be written as
\begin{equation}\label{n5.3}
ds^2=-B(dt+a_\varphi d\tilde{\varphi})^2+ {\sin^2\theta \over B}
d\tilde{\varphi}^2+d\theta^2,
\end{equation}
where $\tilde{\varphi}=\varphi-\Omega_0t$ and
\begin{equation}\label{n5.4}
B=1-\Omega^2_0\sin^2\theta,~a_\varphi=\Omega_0\sin^2\theta B^{-1}.
\end{equation}
The spectrum of frequencies in a non-rotating frame can be easily
found, $\omega_n=n+1/2$, where $n=0,1,..$ is the angular momentum
of a particle on $S^2$. The frequencies in the rotating frame
are $\tilde{\omega}_{nl}=\omega_n+\Omega_0 l$ (see (\ref{1.2})),
where $|l|\leq n$. Thus, the pseudo-trace $K(t)$ is defined as
\begin{equation}\label{n5.5}
K(t)=\sum_{n=0}^\infty\sum_{l=-n}^{n}e^{-\tilde{\omega}_{nl}^2t}.
\end{equation}
The first spectral coefficients in the asymptotic expansion
(\ref{n4.5}) for (\ref{n5.5}) can be found by a direct
computation\footnote{For instance,
one can define for this purpose the
zeta function $\zeta(\nu)=\sum_{nl} \tilde{\omega}_{nl}^{-2\nu}$
and get $c_0$ as the limit 
$2\pi (\nu-1/2)\zeta(\nu)$ at $\nu=1/2$ and
$c_1$ as a limit $-4\pi(\nu+1/2)\zeta(\nu)$ 
at $\nu=-1/2$. These limits can 
be computed directly from the series.} 
\begin{equation}\label{n5.6}
c_0={4\pi \over 1-\Omega_0^2}~,~~c_1={\pi \over 3}\left[2-{1 \over
1-\Omega_0^2}\right].
\end{equation}
We show now that results (\ref{n5.6}) are reproduced by formula
(\ref{2.18}) and $c_0$, $c_1$ can be represented as integrals of
local quantities. To apply (\ref{2.18}) we have to write the
fiducial Hamiltonian in the form (\ref{2.11}) by using the conformal
transformation (\ref{2.8}). Because the theory is conformally
invariant we get\footnote{To get (\ref{n5.7}) one has to take into account
that scalar curvature of 3D space with metric $B^{-1}ds^2$
where $ds^2$ is metric (\ref{n5.2}) is 
$\bar{R}+{1 \over 4} F^{ik}F_{ik}$.}
\begin{equation}\label{n5.7}
\bar{H}^2(\lambda)=-\bar{h}^{ij}(\bar{\nabla}_i+i\lambda a_i)
(\bar{\nabla}_j+i\lambda a_j)
+\frac 18 \bar{R}+{1 \over 32} F^{ik}F_{ik},
\end{equation}
where $a_idx^i=a_\varphi d\tilde{\varphi}$, $\bar{\nabla}_i$ and
$\bar{R}$ are covariant derivatives and the scalar curvature on
the 2-dimensional space $\bar{\cal B}$ with the metric (see
(\ref{n5.3}))
\begin{equation}\label{n5.8}
dl^2={\sin^2\theta \over B^2} d\tilde{\varphi}^2+{1 \over B}
d\theta^2.
\end{equation}
$B$ and $a_\varphi$ are defined in (\ref{n5.4}). For the operator
(\ref{n5.7}) we get
\begin{equation}\label{n5.9}
c_0=\int_{\bar{B}}{\bar h}^{1/2} d^2x=\mbox{Vol}~\bar{\cal B},
\end{equation}
\begin{equation}\label{n5.10}
c_1=\int_{\bar{B}}{\bar h}^{1/2} d^2x \left[{1 \over 24}\bar{R}
-{1 \over 32} \bar{F}^{ij}\bar{F}_{ij}\right],
\end{equation}
where $\bar{F}_{ij}=F_{ij}$ is the Maxwell tensor for the fiducial
potential $a_i$. By using (\ref{n5.4}) one can check that
(\ref{n5.9}) and (\ref{n5.10}) coincide with (\ref{n5.6}). In the
same way one can make the direct check of (\ref{2.18}) for a
conformal field in a rotating 4-dimensional Einstein universe.

\subsection{High temperature limit}
\bigskip

Let us consider a Gibbs state in a stationary space-time
in the limit of high temperatures\footnote{Certainly, 
there must exist necessary physical conditions which
ensure the thermal equilibrium. These conditions require
special analysis in each particular case. We just assume that
in the problems considered here they are satisfied.}.
This limit is one of the most interesting and well
studied regimes. When the parameter $\beta$ is small 
the dominant contribution to the free energy (\ref{i.12}) results
from large frequencies $\omega\simeq \beta^{-1}$. In this
situation one can go from summation in (\ref{i.17}) to the
integral
\begin{equation}\label{n6.1}
F^C(\beta)= \beta^{-1}\int_\mu^{\infty} \Phi(\omega) d\omega \ln
\left(1-e^{-\beta\omega}\right),
\end{equation}
where the spectral density $\Phi(\omega)$ should be determined by
its asymptotic expansion (\ref{2.17}). 
This gives
\begin{equation}\label{2.21}
F^C(\beta)\sim - {1 \over \pi^{D/2}\beta^D}
 \sum_{n=0}^{\infty} \Gamma\left({D-2n \over
2}\right) \zeta(D-2n) c_n\left({\beta \over 2}\right)^{2n}.
\end{equation}
Here $\zeta(x)$ is the Riemann zeta-function. High temperature
asymptotics of this form have been obtained and studied 
on static space-times by Dowker and collaborators \cite{DoKe:78},
\cite{DoSc:88}, \cite{DoSc:89} (see also \cite{NaFu:85}).
Equation (\ref{2.21}) is the extension of these results to non-static
case.

The following remarks regarding (\ref{2.21}) are in order. First,
formula (\ref{2.21}) is purely local. However, the method we
used to get (\ref{2.21}) does not take into account a non-local
term proportional to $\beta^{-1}$, see \cite{DoKe:78},
\cite{DoSc:88}. This contribution
comes for Bose fields
from the region of small $\omega$ where the
logarithm under the integral in  (\ref{n6.1}) results in a term
$\ln (\omega)$. Second, (\ref{2.21}) is obtained 
with the help of dimensional
regularization. When the parameter $D$ coincides with the physical
dimension one of the terms in (\ref{2.21}) has a simple pole.
However if the system is in a finite box or there is a mass gap
$\mu>0$
this singularity is
an artifact of the method. Careful analysis of the infrared
singularities has been carried out recently
by Gusev and Zelnikov \cite{GZ}
who have
derived an approximate non-local expression for 
the one-loop free
energy by using covariant perturbation theory of 
Ref. \cite{BV}.
Further discussion of this problem can be found in \cite{GZ}.

In four dimensions ($D=4$) 
\begin{equation}\label{5.5}
F^C(\beta) \sim -{\pi^2 \over 90} {c_0 \over \beta^4}- {1 \over
24} {c_1 \over \beta^2}-{1 \over 16 \pi^2}\ln(\beta\rho)c_2
-{1 \over 16\pi^{5/2}}\sum_{n=3}^{\infty} \Gamma\left(n-\frac 32\right)
\zeta(2n-3)c_n\left({\beta \over 2\pi}\right)^{2n-4},
\end{equation}
where $\rho$ is a dimensional parameter related to the
regularization and the pole term is omitted. When the frame does
not rotate (\ref{5.5}) coincides with the result
of Ref. \cite{DoKe:78}. 
Of special interest are the leading terms
in (\ref{5.5})
\begin{equation}\label{12.1}
F^C(\beta)\sim -\int d^3x \sqrt{-g}\left[{\pi^2 \over 90}  T^4
+ {1 \over 24} T^2\left(\frac 16 R-V-\frac 23 \Omega^2
\right)\right]-{c_2 \over 16 \pi^2}\ln(\beta\rho).
\end{equation}
Here $T$ is the local temperature $T=1/(\beta\sqrt{B})$ and $R$ is
the scalar curvature of the background space-time. The second term
in r.h.s of (\ref{12.1}) depends on the local angular velocity
$\Omega$ of the system with respect to a local Lorentz frame, see
(\ref{1a.11b}). This result follows from the form of the $c_1$
coefficient (\ref{xx1}) which can be rewritten in terms of the
characteristics of the physical space-time as \cite{F:01}
\begin{equation}\label{11.5}
c_1=\int \sqrt{-g}d^3x {1 \over B}\left[\frac 16
R-V-\frac 23 \Omega^2\right].
\end{equation}
As for $c_2$ in (\ref{12.1}),
we will show in section 3.5 that it does not depend on
the rotation and has a standard covariant form
quadratic in four-dimensional curvatures.

The validity of (\ref{12.1}) can be checked for some
simple models where the spectrum is known. The simplest case is
again conformal fields in rotating Einstein universe. Such models
have been recently investigated by several authors
\cite{HHT}--\cite{LL}. The formula
(\ref{12.1}) is in agreement with
these results.

\subsection{Thermal part of the stress-energy tensor}
\bigskip

The free energy $F^C(\beta)$ of a Gibbs state on a stationary
space-time is a functional of the background metric $g_{\mu\nu}$
and the Killing vector $\xi$. It
is invariant with respect to time independent
coordinate transformations
because these transformations do not change
the form of the metric, see (\ref{1a.7})--(\ref{1a.9}). Infinitesimal
changes of coordinates can be written as
\begin{equation}\label{n7.1}
\delta x^\mu=\zeta^\mu(x^i),
\end{equation}
where $\mu$ runs from $1$ to $D$ and $i$ from $1$ to $D-1$. The
condition that coordinate changes are time independent can written
as
\begin{equation}\label{n7.2}
({\cal L}_\xi \zeta)^\mu=-({\cal L}_\zeta \xi)^\mu=0,
\end{equation}
where $\xi=\partial_t$ is the corresponding Killing field, and
${\cal L}_\xi$ is the Lie derivative along $\xi$. Thus,
(\ref{n7.2}) is equivalent to the condition that the Killing field is
left unchanged under transformations generated by 
$\zeta^\mu(x^i)$. In a stationary space-time the group of
transformations is $D$-dimensional and by the Noether theorem the
invariance of $F^C(\beta)$ implies existence of $D$ Noether currents
and corresponding charges. 
Like in a classical theory, the Noether current
can be defined as $T_{\mu\nu}\xi^{\nu}$
where $T_{\mu\nu}$ is the stress-energy tensor
\begin{equation}\label{n7.3}
T^{\mu\nu}=-{2 \over \sqrt{-g}}\left({\delta F^C(\beta) \over \delta
g_{\mu\nu}}\right)_{\xi}.
\end{equation}
According to (\ref{n7.2}) the variations are to be taken
at fixed $\xi$ \footnote{In practice one has to keep fixed
contravariant components $\xi^\mu$ of the Killing field. 
For instance, variation of
$B=-\xi^2$ then looks as $(\delta B)_\xi=-\xi^\mu\xi^\nu \delta
g_{\mu\nu}$.}. 
Note that $F(\beta)$ is a functional in a $D-1$
dimensional space, so $g_{\mu\nu}$ are functions of $D-1$
coordinates. One can use (\ref{12.1}) to find 
$T^{\mu\nu}$ in
high-temperature limit
As an illustration, we present the result for
conformally coupled scalar field ($V=R/6$ in (\ref{12.1}))
$$
T_{\mu\nu}\sim(g_{\mu\nu}+4u_\mu u_\nu) \left({\pi^2 \over 90}
T^4-{1 \over 36} T^2\Omega^2\right)
$$
\begin{equation}\label{n7.4} 
+{T^2 \over 36}\left(2A_{\nu\rho}A_{\mu}^{~~\rho}-u_\nu
A_{\mu\lambda} w^\lambda-u_\mu A_{\nu\lambda}w^\lambda \right.
\left. +u_\nu A_{\mu\lambda}^{~~~;\lambda}+u_\mu
A_{\nu\lambda}^{~~~;\lambda} \right)+O(\ln T).
\end{equation}
Here $u^\mu$ is the velocity vector of the Killing frame
where thermal equilibrium is defined, $w_\mu$ is its acceleration and
$A_{\mu\nu}$ is the antisymmetric rotation tensor 
(\ref{1a.5}). The invariance of $F(\beta)$ implies the standard
covariant conservation law $T^{\mu\nu}_{~~;\nu}=0$. Indeed, by
considering $F(\beta)$ as local functional (\ref{12.1}) one can
always write the change of $F^C(\beta)$ under coordinate
transformations as
\begin{equation}\label{n7.5}
\delta_\zeta F^C(\beta)
=
\int d^3x \sqrt{-g}\left[\nabla_\mu j^\mu(\zeta,\delta_\zeta
g)-T^{\mu\nu}\zeta_{\mu;\nu}\right],
\end{equation}
where $j_\mu$ is a local expression of $\zeta$ and variations of the
metric. Because the background theory is stationary the divergence
of $j^\mu$ reduces to three-dimensional divergence which results
in a surface term. For stationary transformations (\ref{n7.1}) the
expression (\ref{n7.5}) vanishes which is possible only when
$T^{\mu\nu}$ is divergence free. This justifies its interpretation
as a stress-energy tensor.

By taking into account that $u_\mu A^{\mu\nu}_{~~;\nu}=
-A_{\mu\nu}A^{\mu\nu}$ one can check that $T^\mu_\mu=0$. Hence
(\ref{n7.4}) is interpreted as a stress-energy tensor of a
massless conformally invariant matter with temperature $T$ and 
velocity $u_\mu$. The first term of (\ref{n7.4}) coincides with
the stress-tensor of an ideal gas at temperature $T$.

It should be noted that (\ref{n7.3}) represents the average value
of the thermal part of a
quantum stress tensor-energy tensor in the Gibbs state. It does
not coincide, however, with the total stress-energy tensor because
(\ref{n7.3}) does not
include the vacuum part\footnote{Let us recall that the 
contribution of zero-point energies in
$F^C(\beta)$ has been 
eliminated from the very beginning by normal
ordering, see section 1.1.}.  Because the conformal anomaly
is related to renormalization only of the vacuum energy
this fact explains why  (\ref{n7.4}) is traceless.

High-temperature asymptotics like
(\ref{n7.4}) may represent an interest in studying quantum
effects near rotating black holes because the local temperature $T$
becomes large near the black hole horizon.
To realize such a state for a Kerr--Newman black
hole one has to place it in a cavity which 
co-rotates rigidly with the black hole. Hence, for Kerr-Newman 
black 
hole (\ref{n7.4}) requires to take into account effects of the cavity.  
The situation is simpler for rotating black holes with
asymptotic anti-de Sitter geometry (Kerr--AdS black holes
\cite{Carter:68}). They are solutions in gravity theory with 
negative cosmological constant. If the 
angular velocity of Kerr-AdS black hole is less than inverse
AdS radius the Killing vector field $\xi$ is globally time-like
outside the horizon. 
The cavity is
not necessary in this case and (\ref{n7.4}) may be a correct
approximation to the thermal part of stress-energy tensor, at
least near the horizon.

\bigskip

\section{Euclidean theory}
\setcounter{equation}0

\subsection{Formulation of the problem}
\bigskip

Let us consider a stationary space-time $\cal M$ with metric
defined by (\ref{1a.7})--(\ref{1a.9}). Suppose that the metric depends
on a number of parameters $J$ denoted by a collective symbol $J$.
These parameters will play the same role as the angular velocity
in the metric 
(\ref{1.1}). We assume that there is a complexification
of $\cal M$ (in the sense described in section 1.2) which has a real
Euclidean section ${\cal M}_E$ with metric
\begin{equation}\label{n.i6}
ds_E^2 =
\breve{B}(d\tau+\breve{a}_kdx^k)^2+\breve{h}_{kj}dx^kdx^j.
\end{equation}
$\tau$ is a periodic coordinate with period $\beta$. We also
assume that (\ref{n.i6}) can be obtained from (\ref{1a.7}) by
replacement $t=-i\tau$ and analytical continuation of the
components to imaginary values of $J$
$$
\breve{h}_{kj}(x;J)= h_{kj}(x;iJ)~,
$$
\begin{equation}\label{n.i7}
\breve{a}_{k}(x;J)= ia_{k}(x;iJ)~,~ \breve{B}(x;J)= B(x;iJ)~.
\end{equation}
An example of this procedure has been discussed in section 1.2.
The Euclidean effective action $W^E(\beta,J)$ for free fields or
in one-loop approximation is a determinant of the corresponding
Euclidean operators $L_E$, see (\ref{i.13}). In what follows
we consider the action defined with the help of
$\zeta$ function regularization
\cite{DCH},\cite{Elizalde} as
\begin{equation}\label{n.i8} 
W^E(\beta,J)
=-\frac 12 \lim_{\nu\rightarrow 0}{d \over d\nu}
\zeta(\nu|\beta,J),
\end{equation}
\begin{equation}\label{n.i9}
\zeta(\nu|\beta,J)=\sum_{\Lambda}
\left[\varrho^2\Lambda\right]^{-\nu},
\end{equation}
where $\Lambda$ are eigen-values of $L_E$,
$\varrho$ is a dimensional parameter
related to regularization. The series in (\ref{n.ec23}) converge for 
$Re~\nu>D/2-1$. To
get $\zeta$ function for other values of $\nu$ one
considers analytical continuation.
It can be shown that $\zeta(\nu|\beta)$
is a meromorphic function of $\nu$ with simple poles on the real
axis. In particular, this function is
regular near $\nu=0$ so that one can use (\ref{n.i8}).

To go to the physical space-time one has to
analytically continue (\ref{n.i8}) back to some real values of $J$
(these values are to be in an interval corresponding to the
rotation with the velocity smaller than the velocity of light).
This results in a new functional
\begin{equation}\label{n.i10}
\beta F^E(\beta)=W^E(\beta,-iJ).
\end{equation}
We call $F^E(\beta)$ the Euclidean free energy. Our purpose now is
to describe the relation between $F^E(\beta)$ and canonical free
energy $F^C(\beta)$, Eq. (\ref{i.17})

As earlier we consider the scalar model (\ref{4.3}).
Then $\zeta$-function (\ref{n.i9}) 
is determined by the eigen-values of the 
Euclidean wave operator on ${\cal M}_E$ 
\begin{equation}\label{n.kk13} 
(-\nabla^\mu\nabla_\mu+V)\phi_\Lambda=\Lambda \phi_\Lambda.
\end{equation}
The Kaluza--Klein method can be applied to Euclidean
theory as follows. Note that
because of the isometry, $\phi_\Lambda$
are eigen-vectors of the operator $i\partial_\tau$
\begin{equation}\label{n.kk14}
\phi_{\Lambda}(\tau,x)=e^{i\sigma_l\tau} \phi_{\Lambda}(x),
\end{equation}
where $\sigma_l=(2\pi l)/\beta$,  $l=0,\pm 1, \pm 2,...$. For
these modes (\ref{n.kk13}) takes the form
\begin{equation}\label{n.kk15}
(-\breve{g}^{\mu\nu}_E (\breve{\nabla}_\mu+i\sigma_l
\breve{a}_\mu) (\breve{\nabla}_\nu+i\sigma_l \breve{a}_\nu)+V)
\phi_{\Lambda}=\Lambda \phi_{\Lambda}.
\end{equation}
Here the metric and the connections correspond to a fiducial
static Euclidean space-time $\tilde{\cal M}_E$
\begin{equation}\label{n.kk16}
d\breve{s}^2_E=(\breve{g}_E)_{\mu\nu}dx^\mu
dx^\nu=\breve{B}d\tau^2+ \breve{h}_{jk}dx^j dx^k.
\end{equation}
Thus, the Euclidean problem on a stationary background can be
reformulated as a theory on a static background in the presence of
a fiducial gauge connection.  The distinction from the Lorentzian
theory is that the fiducial charges $\sigma_l$ 
(Matsubara frequencies) are quantized
because the Euclidean time coordinate $\tau$ is compact.

Without loss of generality we again suppose that $\breve{B}=1$ in
(\ref{n.kk16}). One can always use conformal transformation to bring
the metric to this form. The result of this transformation is an
anomalous addition to the Euclidean action. This addition is
proportional to $\beta$ and changes only the
vacuum energy. We will discuss this question in section 3.4.
If
$\breve{B}=1$ the eigen-value problem can be written as
\begin{equation}\label{n.kk17}
(\sigma_l^2+\breve{H}^2(\sigma_l))\phi_{\Lambda}
=\Lambda\phi_\Lambda,
\end{equation}
\begin{equation}\label{n.kk18}
\breve{H}^2(\sigma_l)=-\breve{h}^{jk}(\breve{\nabla}_j+ i\sigma_l
\breve{a}_j) (\breve{\nabla}_k+i\sigma_l \breve{a}_k)+V.
\end{equation}
The operators $\breve{H}(\sigma_l)$ are 
analogous to the Lorentzian
Hamiltonian $H(\lambda)$.  $\breve{H}^2(\sigma_l)$ are
positive-definite elliptic operators. 

Let us  consider first systems on a compact space. 
The spectrum of $\breve{H}^2(\sigma_l)$ is discrete
and bounded from below and 
we suppose it is strictly positive, for simplicity.
Denote the corresponding 
eigen-values by $\breve{\omega}^2(\sigma_l)$.  
Given solutions to the problem
\begin{equation}\label{n.kk19}
\breve{H}^2(\sigma_l)\phi^{(\sigma_l)}_{\omega}(x^i)=
\breve{\omega}^2(\sigma_l)
\phi^{(\sigma_l)}_{\omega}(x^i),
\end{equation}
the solution to (\ref{n.kk13}) can be represented as
\begin{equation}\label{n.kk20}
\phi_{\Lambda}(\tau,x)=e^{i\sigma_l\tau}\phi^{(\sigma_l)}_{\omega}(x^i),~~
\Lambda=\sigma_l^2+\breve{\omega}^2(\sigma_l).
\end{equation}
Then $\zeta$-function (\ref{n.i9}) is
\begin{equation}\label{n.ec23}
\zeta(\nu|\beta)=\varrho^{-2\nu}
\sum_{\sigma_l}\sum_\omega
(\sigma_l^2+\breve{\omega}^2(\sigma_l))^{-\nu}~~~,
\end{equation}
where the sum is taken over
all  eigen-values $\breve{\omega}^2(\sigma_l)$ of $\breve{H}^2(\sigma_l)$
(some eigen-values may coincide). 
The Euclidean
effective action is defined with the help of 
(\ref{n.ec23}) by (\ref{n.i8}).

\subsection{From Euclidean to Lorentzian theory}
\bigskip

By analogy with the
Lorentzian theory, where fiducial charges are continuous, 
let us consider a one-parameter
family of operators $\breve{H}^2(\lambda)$
where the parameter $\lambda$
is real. 
For the considered problem  $\breve{H}^2(\lambda)$ are
positive-definite elliptic operators with  
eigen-values $\breve{\omega}^2(\lambda)$.

We make the following additional assumptions:

1) The Wick rotation to Lorentzian theory is determined by analytical 
continuation of a set of parameters $J$ in such a way that under this 
continuation eigen-values of $\breve{H}^2(\lambda)$ transform into 
eigen-values of the Lorentzian operators $H^2(\lambda)$, i.e.  
\begin{equation}\label{corr}
\breve{\omega}^2(i\lambda|-iJ)=\omega^2(\lambda|J).
\end{equation}
Equation (\ref{corr}) should be considered simply as a 
condition on $\breve{\omega}^2$ as functions of 
$\lambda$ and $J$.

2) As in the Lorentzian theory, see (\ref{constra}),
we define functions
$\breve{\chi}(\breve{\omega},\lambda)
=\breve{\omega}^2(\lambda)+\lambda^2$.
For $\lambda=\sigma_l$ they coincide with
eigen-values $\Lambda$ of the Euclidean operartor $L_E$. 
We assume that 
$\breve{\chi}(\omega,z)^{-1}$ 
can be analytically continued in $z$ to the upper complex plane 
where they are meromorphic functions with simple poles.
Note that according to (\ref{corr}),
$\breve{\chi}(\breve{\omega},\lambda)$
transform to $-\chi(\omega,\lambda)$, functions defined in
(\ref{constra}). Thus, after the Wick rotation the poles lie on 
axes $Re~z=0$ and coincide with physical energies.
The requirement that the poles are simple is equivalent to
condition $\breve{\chi}'(\omega,i\omega)\neq 0$, 
where a prime denotes the derivative with respect to $\lambda$.
This is in accord with condition that
physical states have non-zero norm, see (\ref{C.4}).

For simple models one can check that our second assumption about 
$\breve{\chi}(\breve{\omega},\lambda)$ does hold true.
For instance, for a field in a rotating Einstein universe (discussed 
in section 2.4) the eigen values of the Euclidean operator
$L_E=-\nabla^2+R/8$ are 
$\Lambda=(\sigma_k+l\breve{\Omega}_0)^2 +\omega_n^2$
where $\breve{\Omega}_0$ is the Euclidean angular velocity.
Hence, $\breve{\chi}(\breve{\omega},\lambda)=
(\lambda+l\breve{\Omega}_0)^2 +\omega_n^2$ and
under change $\breve{\Omega}_0$ to $-i\Omega_0$ one gets
$\breve{\chi}(\breve{\omega},i\lambda)=
-(\lambda-l\Omega_0)^2 +\omega_n^2$. They are the functions 
whose zeros $\lambda=\pm (\omega_n+l\Omega_0)$
coincide with physical energies. Also, 
$\breve{\chi}'(\omega,i\omega)=2i\omega_n\neq0$.

We now follow \cite{FZ:01} and rewrite the $\zeta$-function  
by using the Cauchy theorem as 
\begin{equation}\label{n.ec1} 
\zeta(\nu|\beta)={\varrho^{-2\nu} \over 2\pi i}
\sum_{\sigma_l}\sum_\omega
\int_C  
{dz \over z-\sigma_l}
(z^2+\breve{\omega}^2(z))^{-\nu}.
\end{equation}
The contour $C$ consists of two parallel lines, $C_+$ and $C_-$,
in the complex
plane. $C_+$ goes from $(i\epsilon+\infty)$
to $(i\epsilon-\infty)$ and $C_-$  goes from
$(-i\epsilon-\infty)$ to
$(-i\epsilon+\infty)$. Here $\epsilon$ is a small positive
parameter such that $\epsilon<\mu$.   
We consider (\ref{n.ec1}) at $Re~\nu> D/2-1$
and assume that $D\geq 2$.
Summation over $\sigma_l$ in (\ref{n.ec1}) can be performed
\begin{equation}\label{n.ec2}
\zeta(\nu|\beta)={\varrho^{-2\nu}\beta \over 4\pi i}
\sum_\omega
\int_C dz \cot \left({\beta z \over 2}
\right) 
(z^2+\breve{\omega}^2(z))^{-\nu}.
\end{equation}
Note that spectrum $\breve{\omega}^2(\lambda)$
has to be symmetric with respect
to change  $\lambda$ to $-\lambda$. Hence,
integrations over $C_+$ and $C_-$ in (\ref{n.ec2}) coincide
and one can write (\ref{n.ec2}) as twice the integral
over $C_+$.
To proceed let us use the identity
$$
\cot\left({\beta z \over 2}\right)=
{2 \over \beta}{d \over dz}
\ln\left(1-e^{i\beta z}\right)-i$$
which enables one to write
\begin{equation}\label{n.ec3a}
\zeta(\nu|\beta)=\beta\zeta_0(\nu)+
\zeta_T(\nu|\beta),
\end{equation}
\begin{equation}\label{n.ec21a}
\zeta_0(\nu)={\varrho^{-2\nu} \over \pi }\sum_\omega
\int_{0}^{\infty}
(x^2+\breve{\omega}^2(x))^{-\nu}dx,
\end{equation}
\begin{equation}\label{n.ec3b}
\zeta_T(\nu|\beta)={\varrho^{-2\nu} \over  \pi i}
\sum_\omega
\int_{C_+} dz
{d \over dz}
\ln\left(1-e^{i\beta z}\right)
(z^2+\breve{\omega}^2(z))^{-\nu}.
\end{equation}
$\zeta_T(\nu|\beta)$ 
represents the purely thermal part which vanishes
at zero temperature because of small positive imaginary part of $z$
in $e^{i\beta z}$. 
This means that the only quantity which can be responsible for
the vacuum energy is
$\zeta_0(\nu)$.
Let us now deform $C_+$ in (\ref{n.ec21a}) so as to make
the integrand exponentially small at large $z$ due to 
the factor $e^{i\beta z}$.
After that we can integrate by parts to get
\begin{equation}\label{n.ec3}
\zeta_T(\nu|\beta)=
\nu{\varrho^{-2\nu} \over \pi i}\sum_\omega
\int_{C_+} dz 
\ln\left(1-e^{i\beta z}\right)
{ \breve{\chi}'(\breve{\omega},z) \over
(z^2+\breve{\omega}^2(z))^{\nu+1}}.
\end{equation}
We can now represent the Euclidean action in the following
form (see (\ref{n.i8}), (\ref{n.ec3a}))
\begin{equation}\label{n.ec6}
W^E(\beta)
=\beta (\breve{F}(\beta)+ \breve{E}_0),
\end{equation}
\begin{equation}\label{n.ec7}
\breve{F}(\beta)=-\frac 12\lim_{\nu\rightarrow 0}{d \over d\nu}
\zeta_T(\nu|\beta),
\end{equation}
\begin{equation}\label{n.ec9}
\breve{E}_0=-\frac 12\lim_{\nu\rightarrow 0}{d \over d\nu}\zeta_0(\nu),
\end{equation}
where $\zeta_T(\nu|\beta)$ and $\zeta_0(\nu)$ are given by 
(\ref{n.ec21a}) and  (\ref{n.ec3}), respectively.
To compute $\breve{F}(\beta)$
we note that $\zeta_T(\nu|\beta)$ has a form 
$\zeta(\nu|\beta)=\nu f(\nu|\beta)$.
To find $f(\nu|\beta)$ at $\nu=0$ 
we  add to
$C_+$ a large semicircle lying in the upper half of the complex plane 
to make a
closed contour. Because of the exponent 
$e^{i\beta z}$ in the logarithm in
(\ref{n.ec3}) the integration over the semicircle vanishes when its
radius goes to infinity. Our assumption guarantees that the function
$f(\nu|\beta)$ is finite at $\nu=0$ and by the Cauchy theorem its value
is determined by the residues at                           
$z=i\breve{\omega}$.  
We get 
\begin{equation}\label{n.ec26}
\breve{F}(\beta)={1 \over \beta}
\sum_{z}
\ln\left(1-e^{-\beta z}\right),
\end{equation}
where $z$ are corresponding
solutions to $\breve{\chi}(\breve{\omega},iz)=0$.
Finally, the first assumption guarantees that after the Wick rotation
(\ref{n.ec26}) coincides with the canonical free energy
\begin{equation}\label{n.ec27}
F^C(\beta)={1 \over \beta}
\sum_\omega
\ln\left(1-e^{-\beta \omega}\right),
\end{equation}
where the sum is taken over all non-negative 
solutions of the equation $\omega^2(z)=z^2$ 
(see (\ref{constr}), (\ref{constra})).
Thus, according to (\ref{n.i10}), (\ref{n.ec6})
the canonical and Euclidean free energies
are related by Eq. (\ref{main}) where $E_0$ 
coincides with $\breve{E}_0$ after the 
Wick rotation.  We will see in section 3.4 that
$\breve{E}_0$ does correspond to the vacuum energy.

\subsection{Continuous spectrum}
\bigskip

It is instructive to discuss what happens in  case
when the spectrum is continuous. The
Euclidean theory offers another look at
the origin of key relations (\ref{10.29}), (\ref{kk12})
which determine the physical spectral density $\Phi(\omega)$.

The Euclidean action is defined  by (\ref{n.i8})
where the $\zeta$ function now has the form
\begin{equation}\label{n.kk22}
\zeta(\nu|\beta)=\varrho^{-2\nu}
\int_{\mu}^{\infty} d\omega \sum_{\sigma_l}
\breve{\Phi}(\omega;\sigma_l)
(\sigma_l^2+\omega^2)^{-\nu}.
\end{equation}
We assume that $\breve{H}^2(\sigma_l)$ have a positive mass gap 
$\mu$ which does not depend on Matsubara frequencies $\sigma_l$.
Similarly to the discrete case, the $\zeta$-function can 
be rewritten by using the Cauchy theorem in (\ref{n.ec3a})
where 
\begin{equation}\label{n1.ec3b}
\zeta_0(\nu)={\varrho^{-2\nu} \over \pi }
\int_{\mu}^{\infty} d\omega \int_{0}^{\infty}
\breve{\Phi}(\omega;x) (x^2+\omega^2)^{-\nu}dx,
\end{equation}
\begin{equation}\label{n1.ec21a}
\zeta_T(\nu|\beta)={\varrho^{-2\nu} \over  \pi i}
\int_{\mu}^{\infty} d\omega \int_{C_+} dz 
{d \over dz}
\ln\left(1-e^{i\beta z}\right)
\breve{\Phi}(\omega;z)
(z^2+\omega^2)^{-\nu},
\end{equation}
and $C_+$ is defined as earlier.
Then the Euclidean action is represented in the form (\ref{n.ec6})
where $\breve{F}(\beta)$ 
and $\breve{E}_0$  are defined with the help of
(\ref{n1.ec3b}) and (\ref{n1.ec21a}), respectively.

To compute $\zeta_T(\nu|\beta)$ we proceed as before
and deform $C_+$ in (\ref{n1.ec21a}) so as to make
the integrand exponentially small at large $z$ due to 
the factor $e^{i\beta z}$
\footnote{We conjecture that  
$\breve{\Phi}(\omega;z)$ itself cannot increase at large $z$ and 
presence of the factor $e^{i\beta z}$ 
is enough to ensure convergence 
of (\ref{n1.ec21a}).}.  
Integration by parts over $z$ then gives
\begin{equation}\label{n1.ec3}
\zeta_T(\nu|\beta)=
{\varrho^{-2\nu} \over \pi i}
\int_{\mu}^{\infty} d\omega \int_{C_+} dz
\ln\left(1-e^{i\beta z}\right)
\left[{2\nu z \breve{\Phi}(\omega;z) \over
(z^2+\omega^2)^{\nu+1}}-
{\partial_z\breve{\Phi}(\omega;z) \over
(z^2+\omega^2)^{\nu}}
\right].
\end{equation}
One can also integrate by parts over $\omega$
the second term in  square brackets\footnote{The boundary terms 
vanish at $\mbox{Re}~\nu>D/2-1$ because
$\partial_z\breve{\Phi}(\omega;z)\sim \omega^{D-4}$ at large $\omega$.} 
\begin{equation}\label{n.ec4}
\zeta_T(\nu|\beta)=
{\nu \varrho^{-2\nu}\over \pi i}
\int_{\mu}^{\infty} d\omega \int_{C_+} dz
{2z \over (z^2+\omega^2)^{\nu+1}}
\ln\left(1-e^{i\beta z}\right)
\breve{\Psi}(\omega;z), 
\end{equation}
\begin{equation}\label{n.ec5}
\breve{\Psi}(\omega;z)=
\breve{\Phi}(\omega;z)-{\omega \over z} \int_\mu^\omega {d \over dz}
\breve{\Phi}(\sigma;z)d\sigma .
\end{equation}
Function $\breve{\Psi}(\omega;z)$
corresponds to auxiliary density $\Psi(\omega,\lambda)$, see (\ref{kk12}).
In the Lorentzian theory $\Psi(\omega,\lambda)$ appears as result
of analysis  of the inner products. It is remarkable that in the
Euclidean theory there appears an analogous function
$\breve{\Psi}(\omega;z)$, although it happens on a completely different
footing.

To compute $\breve{F}(\beta)$
one can use the fact that $\zeta_T(\nu|\beta)$ now has a form 
$\zeta(\nu|\beta)=\nu f(\nu|\beta)$, see (\ref{n.ec4}). 
We follow \cite{FZ:01} and
assume that for any $\omega$ the density 
$\breve{\Psi}(\omega,z)$ can be analytically continued to complex $z$ 
and is an entire function of $z$ in the upper half of the complex 
plane. 
One can then proceed as in section 3.2 
and compute $f(\nu|\beta)$ at $\nu=0$ by using the Cauchy theorem.
The result looks as follows:
\begin{equation}\label{n.ec11}
\breve{F}(\beta)={1 \over \beta}
\int_{\mu}^{\infty} d\omega
\breve{\Phi}(\omega)
\ln\left(1-e^{-\beta \omega}\right),
\end{equation}
\begin{equation}\label{n.ec12}
\breve{\Phi}(\omega)
=\left.\breve{\Psi}(\omega;z)\right|_{z=i\omega}.
\end{equation}
Similarity between $\breve{\Phi}(\omega)$ and physical density 
$\Phi(\omega)$
defined by (\ref{10.29}), (\ref{kk12}) is obvious.  Consider
now the Wick rotation from Euclidean
${\cal M}_E$ to  Lorentzian space-time $\cal M$  determined by analytical
continuation  of a set of the parameters $J$ of the metric.
Our second assumption is that $\breve{\Psi}(\omega;\lambda)$, as a 
function of parameters $J$, can be analytically continued to complex 
$J$ in such a way that the following equality holds:
\begin{equation}\label{n.ec13}
\breve{\Psi}(\omega;i\lambda|-iJ)=\Psi(\omega;\lambda|J)
\end{equation}
for $J$ and $\lambda$ real.
Condition (\ref{n.ec13}) is an analog of (\ref{corr}).
Given equations (\ref{n.ec11})--(\ref{n.ec13})
one concludes that
$\breve{\Phi}(\omega)$ and $\breve{F}(\beta)$ coincide
with the physical density of levels and the free energy, respectively,
\begin{equation}\label{n.ec14a}
\breve{\Phi}(\omega|-iJ)=\Phi(\omega|J),
\end{equation}
\begin{equation}\label{n.ec14}
\breve{F}(\beta|-iJ)=F(\beta|J).
\end{equation}
A rigorous proof of (\ref{n.ec13}) is problematic
because the explicit form of $\Phi(\omega;\lambda)$ is not
known in general. 
One can demonstrate \cite{FZ:01}, however,  that this assumption
is true at least asymptotically in the limit of
high frequencies when
the spectral density becomes
a local functional  of the background geometry 
and the Wick rotation procedure can be easily applied.

\subsection{Vacuum energy}
\bigskip

Let us return to the vacuum part $\breve{E}_0$ of 
the Euclidean action, see (\ref{n.ec9}).
According to (\ref{n.ec21a}),
\begin{equation}\label{n.v3}
\breve{E}_0=-\frac 12\lim_{\nu\rightarrow 0}{d \over d\nu}\zeta_0(\nu),
\end{equation}
\begin{equation}\label{n.v4}
\zeta_0(\nu)=
{1 \over \pi\varrho}
\int_0^\infty dx
\zeta\left(\nu|\varrho^2 (\breve{H}^2(x)+x^2)\right),
\end{equation}
where $\zeta(\nu|{\cal O})$ is the generalized $\zeta$-function
of an operator $\cal O$.
Let us show that after the Wick rotation $\breve{E}_0$ agrees with the
standard definition
of the vacuum energy. 
For this purpose we represent (\ref{n.ec21a}) in the 
form 
\begin{equation}\label{n.v3a}
\zeta_0(\nu)=-{\varrho^{-2\nu} \over 2\pi }\sum_\omega
\int_{C_+}dz
(z^2+\breve{\omega}^2(z))^{-\nu}
e^{i\epsilon z},
\end{equation}
where
a small positive parameter $\epsilon$  is introduced
to regularize 
the integral in the limit of
vanishing $\nu$. By integrating in (\ref{n.v3a}) 
by parts over $z$, and neglecting terms linear in $\epsilon$,
we get
\begin{equation}\label{n.v3b}
\zeta_0(\nu)=-\nu{\varrho^{-2\nu} \over 2\pi }
\sum_\omega
\int_{C_+}dz
{z \breve{\chi}'(\breve{\omega},z) \over
(z^2+\breve{\omega}^2(z))^{\nu+1}}
e^{i\epsilon z}.
\end{equation}
The regularization enables us to replace $C_+$ by a closed 
contour in the upper half of 
complex plane and use the Cauchy theorem
\begin{equation}\label{n.v3c} 
\breve{E}_0 =\frac 12 \sum_{z} z
e^{-z},
\end{equation}
where $z$ are solutions to $\breve{\chi}(\breve{\omega},iz)=0$. 
After the
Wick rotation $\breve{E}_0$ does correspond to
the vacuum energy determined 
with cutoff $1/\epsilon$ in the range of high frequencies. 
 
A  comment regarding  
relations (\ref{n.ec6}) and (\ref{n.v3c})
is in order. These relations are valid when the Killing 
field $\xi$ has a unit norm, i.e., the space-time is 
"ultrastationary" with metric
\begin{equation}\label{n.v14}
ds^2=(d\tau+\breve{a}_kdx^k)^2+\breve{h}_{kj}dx^kdx^j,
\end{equation}
where $\tau$ is periodic with period $\beta$.
One can always use the
conformal  rescaling
$\bar{g}_{\mu\nu}=g_{\mu\nu}/B$ 
to reduce the 
metric $g_{\mu\nu}$, see (\ref{n.i6}), to ultrastationary form
(\ref{n.v14}). 
The (renormalized)
Euclidean action $W^E[g]$ on (\ref{n.i6})
is related to the action $\bar{W}^E[\bar{g}]$ on (\ref{n.v14}) as
\begin{equation}\label{n.v14a}
W^E[g]=\bar{W}^E[\bar{g}]+\beta \Delta[g],
\end{equation}
where $\beta \Delta[g]$ appears as a result of the conformal
anomaly\footnote{The anomalous term is related
to breaking of conformal invariance in the process of
renormalization of the effective action.
The anomaly can be also attributed to 
non invariance of the integration measure in the Euclidean functional 
integral \cite{Fuji}.}. 
It can be shown that
$\Delta[g]$ is a local functional \cite{DoSc:88},\cite{DoSc:89}.  
According to (\ref{n.v14}),
the vacuum energy computed for
an arbitrary stationary space-time
is $E_0[\bar{g}]+\Delta[g]$,
where $E_0[\bar{g}]$ is defined by 
(\ref{n.v3}), (\ref{n.v4}) on 
on (\ref{n.v14})

Equations (\ref{n.v3}), (\ref{n.v4}) hold for discrete and 
continuous spectra.
They enable one to compute the vacuum
energy of fields in arbitrary stationary space-times
by using the $\zeta$-function method and can be used in
different applications.

\subsection{Kaluza-Klein reduction of heat-kernel coefficients}
\bigskip

Let us discuss now the functions $\zeta(\nu|\beta)$, $\zeta_0(\nu)$
and relation 
(\ref{n.ec3a})
(which holds 
on ultrastationary space-times) in more detail.
We begin with
$\zeta(\nu|\beta)=\zeta\left(\nu|\varrho^2 L\right)$
which is a generalized $\zeta$ function of the
Euclidean operator $L_E=-\nabla^2+V$, see (\ref{n.kk13}) and 
(\ref{n.ec23}).  
According to the general theory, $\zeta(\nu|\beta)$ is 
a meromorphic function which has simple poles on the real axis of 
$\nu$.  In the theory with $D$ dimensions the part 
$\zeta^{(p)}(\nu|\beta)$ which includes the poles of $\zeta(\nu|\beta)$ 
can be written as (see, e.g., \cite{BCVZ:96})
\begin{equation}\label{n.v5}
\zeta^{(p)}(\nu|\beta)={2 \over (4\pi)^{D/2} \Gamma(\nu)}
\sum_{n=0}^\infty {\breve{A}_n \over 2\nu+2n-D},
\end{equation}
where $\breve{A}_n$ are the coefficients related to the asymptotic
expansion of the heat kernel\footnote{In
what follows we put $\varrho=1$ for simplicity.
We will also assume that space-time has no boundaries
and hence $n$ in (\ref{n.v6}) is an integer.}
\begin{equation}\label{n.v6}
\mbox{Tr}e^{-tL_E}\sim {1 \over (4\pi t)^{D/2}}
\sum_{n=0}^\infty \breve{A}_n t^n.
\end{equation}
Relation 
(\ref{n.ec3a}) has an important consequence:
the poles of $\zeta(\nu|\beta)$ coincide with poles
of $\beta\zeta_0(\nu)$. This happens simply because
coefficients $\breve{A}_n$ in (\ref{n.v5}) depend on parameter
$\beta$ linearly. 
To investigate the poles of $\zeta_0(\nu)$ we rewrite 
(\ref{n.v4}) as
\begin{equation}\label{n.v7}
\zeta_0(\nu)=
{1 \over \pi \Gamma(\nu)}
\int_0^\infty dx \int_0^\infty dt~ t^{\nu-1}
\mbox{Tr}~e^{-t(\breve{H}^2(x)+x^2)}.
\end{equation}
Our assumption is that the poles of $\zeta_0$ are related
to the behaviour of the integral at small $t$. The argument
is that in this limit  the trace is a local functional.
By following \cite{BCVZ:96} we define the pole part 
$\zeta^{(p)}_0(\nu)$ of $\zeta_0(\nu)$ by (\ref{n.v7}) with integration 
over $t$ taken in the interval $(0,1)$. One can then replace the trace of 
the heat kernel of $\breve{H}^2(x)$ by its asymptotic 
expansion
\begin{equation}\label{n.as3}
\mbox{Tr} e^{-t\breve{H}^2(\lambda)}
\simeq {1 \over (4\pi t)^{(D-1)/2}}
\sum_{n=0}^\infty\breve{a}_n(\lambda) t^n,
\end{equation}
and get
\begin{equation}\label{n.v8}
\zeta^{(p)}_0(\nu)=
{1 \over \pi \Gamma(\nu)}
\int_0^\infty dx \int_0^1 dt ~t^{\nu-1}
e^{-tx^2}
{1 \over (4\pi t)^{(D-1)/2}}\sum_{n=0}^\infty \breve{a}_n(x)
t^n.
\end{equation}
The integral exists at $\mbox{Re}~\nu>(D-1)/2$.
To proceed we note that the heat kernel coefficients are
polynomials analogous to (\ref{2.16}) 
\begin{equation}\label{n.as8}
\breve{a}_n(\lambda)=\sum_{m=0}^{[n/2]}\lambda^{2m}\breve{a}_{2m,n},
\end{equation}
where $\breve{a}_{2m,n}$ are some coefficients. 
It enables one  to integrate    (\ref{n.v8})  over $x$ and then over $t$
\begin{equation}\label{n.v9}
\zeta^{(p)}_0(\nu)=
{2 \over  (4\pi)^{D/2}\Gamma(\nu)}
\sum_{n=0}^\infty\sum_{m=0}^{[n/2]}
{\Gamma(m+1/2)\breve{a}_{2m,n} \over \sqrt{\pi}(2\nu+2(n-m)-D)}.
\end{equation}
The latter equation can be rewritten also as
\begin{equation}\label{n.v10}
\zeta^{(p)}_0(\nu)=
{2 \over  (4\pi)^{D/2}\Gamma(\nu)}
\sum_{n=0}^\infty {1 \over 2\nu+2n-D}
\sum_{m=n}^{2n}
{\Gamma(m-n+1/2) \over \sqrt{\pi}}
\breve{a}_{2(m-n),m}.
\end{equation}
By comparing poles of (\ref{n.v5}) and (\ref{n.v10}) one
concludes that
\begin{equation}\label{n.v11}
\breve{A}_n={\beta  \over \sqrt{\pi}}
\sum_{m=n}^{2n} \Gamma(m-n+1/2)
\breve{a}_{2(m-n),m}.
\end{equation}
For $D$ odd this conclusion is true for all $n$ while for $D$ even only
for $0\leq n \leq D/2-1$ (terms with other $n$
do not result in poles). It is clear, however, that (\ref{n.v11})
should be universal for all $D$ and all $n$ because dimensionality
does not appear in it explicitly.

Equation (\ref{n.v11}) can be considered as a
Kaluza--Klein reduction formula for heat kernel coefficients
of an operator on a stationary $D$ dimensional manifold.
It has one interesting and
important consequence at $n=D/2$
for even $D$ .  Namely, quantity $\beta^{-1}\breve{A}_{D/2}$
transforms under the Wick rotation to the spectral coefficient
$c_{D/2}$ defined by (\ref{2.18}).
This can be easily seen 
with the help of (\ref{n.v11}) if we take into account
the relation between Euclidean and Lorentzian coefficients
\begin{equation}\label{n.as5}
\breve{a}_n(-iJ,i\lambda)=
a_n(J,\lambda).
\end{equation}
Equations (\ref{n.as5}), (\ref{2.16}) and (\ref{n.as8}) imply that
\begin{equation}\label{n.as5a}
\breve{a}_{2m,n}(-iJ)=(-1)^m
a_{2m,n}(J).
\end{equation}
The fact that $c_{D/2}$ is related to $\beta^{-1}\breve{A}_{D/2}$
is remarkable for two reasons. 
First, because
$\breve{A}_{D/2}=(4\pi)^{D/2}\zeta(0|\beta)$  and
$\zeta(0|\beta)$ determines the anomalous
scaling of the Euclidean effective action \cite{BiDa:82}.
Second, because this fact implies 
that $c_{D/2}$ is a local covariant functional 
of $D$-dimensional curvatures. One can, thus, conclude that the
coefficient $c_2$ which appears in high temperature
asymptotic (\ref{12.1}) in four dimensions
is a covariant functional quadratic
in curvatures, the same property it has in static space-times.

The argument we used to derive (\ref{n.v11}) is not
quite rigorous.
Relation (\ref{n.v11}), however, passes a non-trivial 
check.  Consider (\ref{n.v11}) for $n\neq 0$ on a $D$-dimensional 
stationary Euclidean background (\ref{n.v14}) where $\tau$ is periodic 
with period $\beta$.  We denote this space by ${\cal M}_E$. The 
$D-1$-dimensional space $\cal B$ corresponding to operators 
$\breve{H}^2(x)$ is determined by the metric 
\begin{equation}\label{n.v15}
dl^2 =h_{kj}dx^kdx^j.
\end{equation}
We denote the Riemann tensors on ${\cal M}_E$ and $\cal B$
as $\breve{R}_{\mu\nu\lambda\rho}$ and $\bar{R}_{\mu\nu\lambda\rho}$,
respectively. Their relation is discussed in Ref. \cite{FZ:01}.

Consider (\ref{n.v11}) for $n=1$. It
can be shown \cite{FZ:01} that in any dimension
\begin{equation}\label{n.v19}
\breve{A}_1
=\beta \int \sqrt{h}d^{D-1}x\left(\frac 16 \breve{R}-V\right)
=
\beta \int \sqrt{h}d^{D-1}x\left(\frac 16 \bar{R}-{1 \over 24}
F^{\alpha\beta}F_{\alpha\beta}-V\right),
\end{equation}
where the last line holds on ultrastationary space-times.
On the other hand, according to (\ref{n.v11})
\begin{equation}\label{n.v20}
\breve{A}_1=\beta \left(\breve{a}_{0,1}+\frac 12 \breve{a}_{2,2}\right).
\end{equation}
By using heat kernel asymptotics in external gauge field
one finds
\begin{equation}\label{x1}
\breve{a}_{0,1}=
\int \sqrt{h}d^{D-1}x\left(\frac 16 \bar{R}-V\right),
\end{equation}
\begin{equation}\label{x2}
\breve{a}_{2,2}=-
\int \sqrt{h}d^{D-1}x{1 \over 12}F^{\alpha\beta}F_{\alpha\beta}.
\end{equation}
Hence, the right hand sides of (\ref{n.v19}) and (\ref{n.v20})
do coincide.
Equation  (\ref{n.v11}) can be also checked for 
the coefficient $\breve{A}_2$, 
which on a 
manifold without boundary looks as
\begin{equation}\label{n.a12}
\breve{A}_2=
\beta \int \sqrt{h}d^{D-1}x\left(
{1 \over 72} \breve{R}^2-{1 \over 180} \breve{R}_{\mu\nu}
\breve{R}^{\mu\nu}
+{1 \over 180} \breve{R}_{\mu\nu\lambda\rho}\breve{R}^{\mu\nu\lambda\rho} 
-\frac 16 \breve{R}V +\frac 12 V^2
\right)
\end{equation}
The relation (\ref{n.v11}) for $n=2$  is
\begin{equation}\label{a13}
\breve{A}_2=\beta \left(\breve{a}_{0,2}+\frac 12 \breve{a}_{2,3}
+\frac 34 \breve{a}_{4,4}\right).
\end{equation}
The coefficients $\breve{a}_{0,2}$,
$\breve{a}_{2,3}$, $\breve{a}_{4,4}$
can be found by using
results of Refs. \cite{Avramidi}, \cite{BGV}, i.e.
\begin{equation}\label{n.a15}
\breve{a}_{0,2}=
\int \sqrt{h}d^{D-1}x\left(
{1 \over 72} \bar{R}^2-{1 \over 180}
\bar{R}_{\mu\nu}\bar{R}^{\mu\nu} 
+{1 \over 180} \bar{R}_{\mu\nu\lambda\rho}
\bar{R}^{\mu\nu\lambda\rho}-\frac 16 \bar{R}V
+\frac 12 V^2
\right),
\end{equation}
$$
\breve{a}_{2,3}=
\int \sqrt{h} d^{D-1}x\left({1 \over 12} F^{\mu\nu}F_{\mu\nu} V
+{1 \over 60} F_{\mu\rho}~^{||\rho}F^{\mu\lambda}~_{||\lambda}
-{1 \over 72} (F^{\mu\nu}F_{\mu\nu})\bar{R} \right.
$$
\begin{equation}\label{n.a16}
\left.
-{1 \over 180}\bar{R}^{\lambda\nu\mu\rho}F_{\lambda\nu}F_{\mu\rho}
-{1 \over 90}\bar{R}_{\mu\nu}F^{\rho\mu}F_{\rho}~^\nu \right),
\end{equation}
\begin{equation}\label{n.a17}
\breve{a}_{4,4}=\int \sqrt{h} d^{D-1}x \left(
{1 \over 288} (F^{\mu\nu}F_{\mu\nu})^2
+{1 \over 360}
F^{\lambda\nu}F^{\mu\rho}F_{\lambda\mu}F_{\nu\rho}\right).
\end{equation}
Here the symbol $||$ corresponds to covariant
differentiation with respect to the metric $h_{\mu\nu}$.
By using (\ref{a13})--(\ref{n.a17}) one can check that (\ref{n.a12}) 
holds \cite{FZ:01}.

Formulas (\ref{n.v11}) are also interesting in applications
to genuine Kaluza--Klein theories. In this case the Euclidean
space ${\cal M}_E$ has to be identified with a higher dimensional
space, the Euclidean time with an extra coordinate and
$\beta$ with a compactification radius.

\section{Generalization and external gauge fields}
\setcounter{equation}0

As we have seen the Euclidean formulation of 
statistical mechanics offers an alternative derivation
of relations (\ref{10.29}), (\ref{kk12}) between physical and fiducial 
spectral densities.  The advantage of this derivation 
is that actually it does not depend on the form of the Euclidean
wave operator $L_E$. 
Hence, one can conjecture that (\ref{kk12}) is more universal
and may be true (with some additional assumptions) 
when $L_E$ is a higher-order elliptic operator.
If this were the case such  results
of Section 2 as spectral asymptotics
could be extended to non-linear
spectral problems of a generic form.
We will not analyze this possibility here but
mention another problem where (\ref{10.29}) and (\ref{kk12})
are satisfied.
This problem is interesting
for physical applications.   

Consider a charged scalar field in Minkowski space-time
which interacts with a static electric potential $\varphi(x^i)$.
The equation of motion is 
\begin{equation}\label{ne4.1}
\left(-{\cal D}_\mu{\cal D}^\mu+m^2\right)\phi=0,
\end{equation}
where ${\cal D}_\mu=\partial_\mu-igA_\mu$,
$A_\mu dx^\mu=\varphi dt$ and $g$ is the charge of the field. 
For single-particle excitations
$\phi_\omega(t,x^i)=e^{-it\omega}\phi_\omega(x^i)$
is reduced to the non-linear spectral 
problem 
\begin{equation}\label{ne4.2}
\left(\omega^2-H^2(\omega)\right)\phi_\omega(x^i)=0,
\end{equation}
\begin{equation}\label{ne4.3}
H^2(\omega)=-\partial_i^2-g^2\varphi^2-2\omega g\varphi +m^2.
\end{equation}
There must be some physical
restrictions on the potential 
$\varphi$ 
so as to avoid complex energies $\omega$ 
in (\ref{ne4.2}).
Complex $\omega$ correspond to an instability of
the system (creation of particle--anti-particle pairs,
for example). To exclude such effects
we assume that electric field is sufficiently weak
and vanishes at spatial infinity. Note that 
equation (\ref{ne4.1}) preserves its form under the gauge 
transformations $\phi'=e^{-igat}\phi$, $\varphi'=\varphi+a$,
where $a$ is a constant. These transformations shift the spectrum
of $\omega$ in (\ref{ne4.2}) to $\omega+ga$. We will eliminate
this arbitrariness by requiring that $\varphi$ is zero at
infinity.

One can now proceed as before and formulate a 
corresponding fiducial
problem
\begin{equation}\label{ne4.4}
\left(-\partial_\mu\partial^\mu
-g^2\varphi^2-2g\lambda\varphi+ m^2
\right)\phi^{(\lambda)}=0.
\end{equation}
The Klein--Gordon inner product for physical
and fiducial fields are determined, respectively, by
the currents
\begin{equation}\label{ne4.5}
j_\mu(\phi_1,\phi_2)=-i(\phi_1^{*}{\cal D}_\mu\phi_2
-({\cal D}_\mu\phi_1)^{*}\phi_2),
\end{equation}
\begin{equation}\label{ne4.6}
\tilde{j}_\mu(\phi_1,\phi_2)=-
i(\phi_1^{*}\partial_\mu\phi_2
-\partial_\mu\phi_1^{*}\phi_2).
\end{equation}
It should be noted that for charged fields 
one has to consider two Hamiltonians: one is $H(\omega)$
defined in (\ref{ne4.3}) and the other is $H(-\omega)$,
for $\omega>0$.
The reason is that the system contains particles
and antiparticles which have different charges and 
interact with the electric field in different ways.
The particles and antiparticles are described by 
$H(\omega)$ and $H(-\omega)$, respectively.
The total spectral density $\Phi(\omega)$ is a sum of the 
spectral densities of particles, $\Phi_+(\omega)$,
and antiparticles  $\Phi_-(\omega)$.
The fiducial problem for antiparticles has to
be formulated as (\ref{ne4.4}) with $\lambda$ 
replaced by $-\lambda$. Suppose that $\Phi_{\pm}(\omega;\lambda)$
are fiducial spectral densities
for particles and antiparticles  
and $\Psi_{\pm}(\omega;\lambda)$ are corresponding auxiliary 
densities defined by (\ref{kk12}).
Then
by using (\ref{ne4.5}), (\ref{ne4.6}) one can 
prove relation (\ref{10.29}), i.e. that
$\Phi_{\pm}(\omega)=\Psi_{\pm}(\omega;\omega)$.
The asymptotic expansion for the total density 
$\Phi_+(\omega)+\Phi_-(\omega)$ is given by (\ref{2.17}),
(\ref{2.18}).
The spectral coefficients are $c_n=c_{+,n}+c_{-,n}$
where $c_{\pm,n}$ are determined by the heat kernel
asymptotic expansion of operators $H^2(\lambda)$
and $H^2(-\lambda)$. 
These results can be used to show that an electric field
yields a correction $T^2\varphi^2$ to the high-temperature
asymptotics (\ref{12.1}).

One can come to the same result
by considering Euclidean theory. Contributions from
particles and antiparticles in this case
appear in Euclidean $\zeta$-function,
 (\ref{n.ec1}), as integrations 
along contours $C_+$ and $C_-$ lying 
in the upper and lower parts of the complex plane, respectively.
For charged fields these integrations do not 
coincide.

\section{Quantum fields near black holes}
\setcounter{equation}0
\subsection{Motivations}

As we have already discussed, 
quantum fields 
in thermal equilibrium with a black hole near its horizon
appear, to a static observer, as a system at high temperatures. 
The free energy of such fields is described by
high-temperature asymptotic in a form
like (\ref{12.1}). The local temperature $T$ becomes infinite
at the horizon.
To avoid this divergence the integration in (\ref{12.1}) has to be
stopped at  some
small (proper) distance $\it l$ near the horizon.
It is easy to estimate that the entropy of the gas
is of order of ${\cal A}/ {\it l}^2$ where
${\cal A}$ is the surface area of the horizon.
As was first pointed out by 't Hooft \cite{Hooft:85},
if
$\it l$ is associated with the Planck length ${\it l}_{Pl}$
the entropy of the gas is of the same order as the
entropy of a black hole
$S^{BH}={\cal A}/ 4G$ (here $G={\it l}_{Pl}^2$ is the
Newton constant).  

The entropy of black holes
$S^{BH}$ was introduced by Bekenstein \cite{Beke:72-74}
and Hawking \cite{Hawking:75}
who used the fact that black holes possess properties 
similar to the properties of thermodynamical systems
\cite{BCH:73}. In a classical theory, however, 
a Schwarzschild black hole
is nothing but an empty space with a strong gravitational
field. 
It still remains a fundamental problem how
to identify microscopical degrees of freedom of a black  
hole responsible for $S^{BH}$. 
The observation by 't Hooft is important because 
it offers one of the possible explanations of 
the Bekenstein-Hawking entropy.

We do not give here a detailed analysis of this 
idea and do not include a (still growing) list of 
related references.
This can be found in a review paper \cite{FF:98}.
In remaining sections we discuss a single problem, i.e.,
the correspondence between canonical, $F^C$,
and Euclidean, $F^E$, free energies in external regions
of black holes. 
In particular we discuss
the duality property between 
infrared thermal divergencies of $F^C$
and ultraviolet divergencies
of $F^E$ caused by conical singularities.
This is an interesting feature which is
important in finding connection
between $S^{BH}$ and the entropy of a thermal
atmosphere around a black hole.

Note that more detailed analysis of this problem 
is presented in Refs. \cite{F:98},\cite{FF:98}. Our aim here 
is to sketch the main results and pay attention to features 
related to rotation of a black hole.

\subsection{Killing horizons}
\bigskip

We begin with definitions.
Consider a space-time $\cal M$ which possesses a one-parameter
group of isometries parametrized by $t$ and generated
by a Killing vector field $\xi^\mu$. Suppose the isometries
leave fixed each point of a smooth space-like two-surface $\Sigma$.
The surface $\Sigma$ is called a fixed point set of $\xi^\mu$.
It can be shown \cite{KaWa:91} that existence of $\Sigma$
implies the existence of a Killing horizon on $\cal M$.
The Killing horizon consists of two hypersurfaces
spanned by null geodesics and the Killing field is tangent to 
these geodesics.
The two hypersurfaces orthogonally intersect at $\Sigma$, and that is 
why $\Sigma$ is a bifurcation surface for orbits generated by $\xi$.  
The structure of these orbits near $\Sigma$ is similar to 
orbits of Lorentz boosts in Minkowski space-time.

The event horizon of black hole solutions
in Einstein theory is the  Killing horizon\footnote{Note,
however, that we do not
require in this Section that the metric obeys the Einstein equations.}.
The Killing horizon divides the space-time into four regions:
two regions $\cal F$ and $\cal P$ where $\xi$ is space-like, 
and two causally independent 
regions $\cal L$ and $\cal R$ where $\xi$ is time-like.  
In case of black hole space-times one of such 
regions, say $\cal R$, describes 
the exterior region  of a black hole.
In such a region near $\Sigma$ the metric can be written 
in the form 
\begin{equation}\label{na6.1}
ds^2=-B(dt+a_\alpha dx^\alpha)^2+d\rho^2+\sigma_{\alpha\beta}
dx^\alpha dx^\beta,
\end{equation}
where $x^\alpha$ is a set of $D-2$ coordinates and
$\Sigma$  is the surface $\rho=0$. The coordinate $\rho$
is the distance 
measured along a geodesic normal to $\Sigma$
between a point with coordinates $\rho,x^\alpha$
and a point on $\Sigma$ with coordinates $x^\alpha$.
The components of (\ref{na6.1}) do not depend on $t$.
One can show that 
$B\simeq \kappa^2\rho^2$ at small $\rho$.  
Here $\kappa$ is the surface gravity, a constant 
which can be also defined in a covariant way 
in terms of the Killing field
\begin{equation}\label{n6.1a}
\kappa=\left[-\frac 12 \xi_{\mu;\nu}\xi^{\mu;\nu}
\right]_{\xi^2=0}.
\end{equation}

The Euclidean space ${\cal M}^E$ corresponding to $\cal M$ 
is obtained by the Wick rotation as we described earlier.
The analog of $\Sigma$ on ${\cal M}^E$ is a hypersurface
which is a fixed-point set of the Euclidean Killing vector field
$\xi^E$. The Euclidean metric can be written in a form
similar to (\ref{na6.1}) 
where $t$ is replaced by $-i\tau$.
One can easily see that 
there are conical singularities on $\Sigma$ if the period of Euclidean
time $\tau$ 
is arbitrary. ${\cal M}^E$ is regular if the period is
$2\pi /\kappa$. 

The geometry near Lorentzian or Euclidean horizons
can be characterized by geometric invariants obtained
by projecting components of the Riemann tensor and its 
derivatives on 
directions orthogonal to $\Sigma$. The corresponding projector
is defined in terms of two vectors $l$ and $p$ normal to $\Sigma$
as
$P^{\mu\nu}=l^\mu l^\nu \mp p_\mu p_\nu$, where 
$l^2=1$, $p^2=\mp 1$, $(l\cdot p)=0$. Signes $-$ and $+$
correspond to Lorentzian and Euclidean theoriues respectively.

\subsection{Canonical free energy}
\bigskip

In the presence of a
Killing horizon, the spectrum of single-particle
energies $\omega$ has a number of specific features. 
To better understand them let us first
return to the high-temperature asymptotic (\ref{12.1}).
The only difference between stationary and static space-times
in this formula 
is the presence of the term $T^2\Omega^2$ which appears
when the Killing frame rotates with the local angular velocity
$\Omega$. By using the
definition of $\Omega$ (see (\ref{1a.11b}))
and the asymptotic form $B\sim \kappa^2 \rho^2$ 
in (\ref{na6.1}) it can be easily shown that
$\Omega^2\sim \rho^2$ near the horizon and, hence,
the term $T^2\Omega^2$ is finite at $\rho=0$. This means 
that the contribution of rotation in (\ref{12.1})  
can be neglected as compared to other terms
(proportional to $T^4$ and $T^2$). 
Hence, fiducial potential $a_\alpha$ 
in (\ref{n6.1}) can be ignored and
stationary space-time
$\cal M$, Eqs. (\ref{1a.7})--(\ref{1a.9}), can be replaced  by a 
fiducial static space-time $\tilde{\cal M}$, Eq. (\ref{4.10}).
Really, {\it near the horizon one has effectively a static
problem} \cite{FF:00}. 

The last statement can be also formulated
in geometrical terms.
Consider three invariants
$P^{\mu\nu}P^{\lambda\rho}R_{\mu\lambda\nu\rho}$,
$P^{\mu\nu}R_{\mu\nu}$, and $R$ computed at $\Sigma$,
where $R_{\mu\nu\lambda\rho}$, $R_{\mu\nu}$ and
$R$ are $D$--dimensional Riemann, Ricci and scalar curvatures,
respectively. It can be shown \cite{FF:00} that 
these quantities computed on $\cal M$
coincide with corresponding quantities computed on $\tilde{\cal M}$.
Let us also note that surface gravities $\kappa$ of the Killing horizons
as well as the bifurcation surfaces 
on $\cal M$ and $\tilde{\cal M}$ are identical.

To understand what
happens near the horizon let us neglect for a moment the
curvatures near $\Sigma$
and approximate the
metric by the metric on the Rindler space 
\begin{equation}\label{rind}
ds^2= -\kappa^2\rho^2 dt^2+d\rho^2+dz_1^2+...+dz_{D-2}^2.
\end{equation}
As was explained in section 2.2. 
the single-particle Hamiltonian can be obtained 
with the help of a conformal
transformation which turns
(\ref{rind}) into an ultrastatic space. From (\ref{2.11}), (\ref{2.13})
one finds   
\begin{equation}\label{H}
H^2=-\nabla_i\nabla^i-\alpha_D^2\kappa^2+\kappa^2 \rho^2 V.
\end{equation}
Here $\nabla_i$ are covariant derivatives on the
space $\bar{\cal B}$ with 
the metric
\begin{equation}\label{2.4.3}
dl^2=\kappa^{-2}\rho^{-2}(d\rho^2+dz_1^2+...+dz_{D-2}^2).
\end{equation}
$\bar{\cal B}$ is a hyperbolic (Lobachevsky) space with
constant negative curvature $\bar{R}=-\kappa^2(D-1)(D-2)$. 
Although to leading approximation 
the potential $V$ in (\ref{H})
can be neglected,
$H^2$ has an effective tachyonic mass $-\alpha_D^2\kappa^2$
where $\alpha_D=(D-2)/2$.
In its turn,
$-\nabla^i\nabla_i$ on $\bar {\cal B}$
has a mass gap which appears due to
constant curvature.  
What happens is that the tachyonic mass exactly
cancels the mass gap (see \cite{CVZ:95a},\cite{BCZ:96} 
for details). Thus, the spectrum of $H^2$ is continuous
and without mass gap.

Because (\ref{2.4.3}) is a maximally symmetric space
the spectral
density of $H^2$ grows as the volume
of $\bar{\cal B}$. The volume divergence is of the infrared
type and one can show that it appears in 
high-temperature asymptotics (\ref{12.1})
as divergence of the $T^4$ term.

In previous sections our approach to volume divergences
was to cut off integrations over the space to have a system
in a box.
In case of black holes
it means that  the region of  physical space-time whose 
proper distance $\rho$ from the horizon is smaller than some length
$\epsilon$ should not be considered
\cite{FrNo:93}.  
The drawback of this procedure is that
it makes the space-time incomplete. Thus, it makes sense
to consider 
other
regularizations which allow one to work on an entire 
external region of a black hole. The dimensional \cite{F:98}
and Pauli--Villars \cite{DLM:95} regularizations are typical
examples of this type.

We begin with the dimensional regularization.
The idea is that 
the volume of (\ref{2.4.3})
depends on the number of dimensions $D-1$ and
it diverges as
$\epsilon^{2-D}$ where $\epsilon$ is the cutoff at small $\rho$
($\rho>\epsilon$).
Formally, if $D< 2$, the integral at small $\rho$ converges
so one can use $D$ as a
regularization parameter. 
Consider as an example a massive scalar 
field\footnote{The mass term eliminates  
divergencies at large $\rho$.} 
described by Eq.
(\ref{4.3}) with $V=m^2$. Near the horizon 
the diagonal element of the trace of the heat kernel
of (\ref{H}) behaves as \cite{F:98}
\begin{equation}\label{n6.2a}
\left[ e^{-H^2t} \right]_{\mbox{diag}}
\sim
{e^{-m^2\kappa^2\rho^2 t}
 \over (4\pi t)^{(D-1)/2}}
\left(1+b_1t+b_2t^2+...\right),
\end{equation}
\begin{equation}\label{n6.3a}
b_1\sim \kappa^2\rho^2
\left(\frac 16-\xi\right) R,~~
b_2=O(\rho^4).
\end{equation}
Other terms in $b_1$ do not 
result in singularities at
$D\rightarrow 4$.
The integration in the trace near $\rho=0$ can be written as
\begin{equation}\label{n6.5a}
\int_{\bar{B}}\sqrt{\bar{g}}d^{D-1}x
\sim
{1 \over \kappa^{D-1}}
\int_{\Sigma}\int_{0}
\rho^{1-D}d\rho\left[1+\frac 14
\rho^2 {\cal P}
\right],
\end{equation}
where ${\cal P}$ is the curvature invariant
computed on $\Sigma$
\begin{equation}\label{n6.2} 
{\cal P}= 
2P^{\mu\nu}P^{\lambda\rho}R_{\mu\lambda\nu\rho}
-P^{\mu\nu}R_{\mu\nu}.
\end{equation}
In case of
stationary but not static space-times
$\cal P$ has to be computed by using the curvatures
of the corresponding fiducial static background $\tilde{\cal M}$.
However, 
as we have mentioned,  
curvatures $\cal P$ on $\tilde{\cal M}$ and $\cal M$
coincide \cite{FF:00}. 

Equations (\ref{n6.2a}), (\ref{n6.5a}) can be used
to derive the divergent part of the trace 
\begin{equation}\label{n6.6a}
\left[\mbox{Tr}~
e^{-H^2t}\right]_{\mbox{div}}=
{\Gamma\left(1-\frac D2\right) \over (4\pi)^{(D-1)/2}}
 {m^{D-4} \over 2\kappa t^{3/2}} 
\int_{\Sigma}
\left[\left(m^2-\left(\frac 16 -\xi\right)R\right)t
-{{\cal P} \over 4\kappa^2}\right].
\end{equation}
From (\ref{n6.6a}) one can also get the spectral density 
of $H^2$ in dimensional regularization (see (\ref{11.1}))
\begin{equation}\label{n6.7a}
\Phi_{\mbox{div}}
(\omega,D)=
{\Gamma\left(1-\frac D2\right) \over (4\pi)^{D/2}}\,\,
 {m^{D-4} \over \kappa}
\int_{\Sigma}
\left[2\left(m^2-\left(\frac 16 -\xi\right)R\right)
-{\omega^2\over \kappa^2}{\cal P}\right],
\end{equation}
and from it the free energy
\begin{equation}\label{n6.8a}
F^C_{\mbox{div}}(\beta,D)=-
{\Gamma\left(1-\frac D2\right) \over (4\pi)^{D/2}}
\,\,{\pi^2 m^{D-4} \over 3\kappa \beta^2}
\int_{\Sigma}
\left[m^2- \left({1 \over 60}
{4\pi^2 \over \kappa^2\beta^2} {\cal P}+
\left(\frac 16-\xi\right) R\right)
\right].
\end{equation}
The divergencies in (\ref{n6.7a}), (\ref{n6.8a})
appear as poles at $D=4$. This is a
typical property of dimensional regularization.

Expression (\ref{n6.7a}) is suitable
to go to  the Pauli--Villars (PV) regularization.
Let us introduce  $5$ 
additional 
auxiliary scalar
fields (PV partners): $2$  with
masses $M_k$ and  $3$
with masses $M_r'$, and impose two
restrictions 
\begin{equation}\label{n6.9a}
f(1)=f(2)=0,
\end{equation}
where
\begin{equation}\label{3.4ab}
f(p)=m^{2p}+\sum_k M_k^{2p}-\sum_r(M'_r)^{2p}=0.
\end{equation}
Equation
(\ref{n6.9a}) has the solutions
$M_{1,2}=\sqrt{3\mu^2+m^2}$, $M'_{1,2}=\sqrt{\mu^2+m^2}$,
$M'_3=\sqrt{4\mu^2+m^2}$ where $\mu$ is a regularization
parameter,
see \cite{DLM:95}.
The regularized density of states in PV method
is defined as
\begin{equation}\label{n6.10a}
\Phi(\omega,\mu)
\equiv\Phi(\omega,m)+\sum_k 
\Phi(\omega,M_k) 
-\sum_r \Phi(\omega,M'_r),
\end{equation}
where $\Phi(\omega,M_k)$ and 
$\Phi(\omega,M_r')$ are spectral densities of 
PV partners. Fields with masses $M'_r$ give
negative contributions to the regularized density
(as if these scalars had a wrong, Fermi, statistics).

Suppose now that we start with dimensionally regularized 
theory. At $D\neq 4$ the spectral density for each PV
partner is  a finite quantity 
given by (\ref{n6.7a}).
Although each of these densities has a pole the 
poles are cancelled in the total 
density (\ref{n6.10a}) by virtue of the
constraints (\ref{n6.9a}). Hence, one can put 
in (\ref{n6.10a}) $D=4$ and get in the limit $\mu\gg m$
\begin{equation}\label{n6.11a}
\Phi_{\mbox{div}}
(\omega ,\mu)=
{1 \over (4\pi)^2 \kappa}
\int_{\Sigma}
\left[2c\mu^2
+\ln{\mu^2 \over m^2}
\left({\omega^2 \over \kappa^2}{\cal P}+
2\left(\frac 16 -\xi\right)R-2m^2\right)\right],
\end{equation}
where $c=\ln(729/256)$.
The corresponding expression for the divergent part of
canonical free energy is
\begin{equation}\label{n6.12a}
F^C_{\mbox{div}}(\beta,\mu)=-
{1 \over 48\kappa \beta^2}
\int_{\Sigma}
\left[2c\mu^2
+\ln{\mu^2 \over m^2}
\left({1 \over 60}
{4\pi^2 \over \kappa^2\beta^2} {\cal P} +
\left(\frac 16-\xi\right) R-m^2\right)
\right].
\end{equation}
The Pauli-Villars regularization  
effectively results in a cutoff of the
integrals near the horizon at the proper distance comparable to the
inverse  mass of PV partners.
This fact allows the following interpretation. 
Near the horizon, where the local temperature becomes greater than 
$\mu$, PV fields become thermally excited and by virtue of the
constraints (\ref{n6.9a}) their contribution exactly cancels the 
contribution of the physical field.

Usually dimensional and Pauli--Villars regularizations are
used to regularize the ultraviolet divergencies.
As we have seen, in case of black holes they can be also used
to avoid infrared singularities near the horizon.
We will show now that this is not an accident and
the divergences of canonical free energy in these regularizations 
match precisely ultraviolet divergences of 
Euclidean free energy.

\subsection{Heat kernel expansion and conical
singularities}
\bigskip

Let us denote by ${\cal M}^E_\beta$ a Euclidean manifold
which possesses a one-parameter group of isometries.
Suppose that $\Sigma$ is a fixed-point set of these isometries
and that ${\cal M}^E_\beta$ has a structure
$C_{\kappa\beta} \times R^{D-2}$ near $\Sigma$.
Here $C_\alpha$ is a two-dimensional
conical space (the value $\alpha=2\pi$ corresponds to
a plane).

One has the following heat kernel expansion 
of a Laplace operator on ${\cal M}^E_\beta$
\begin{equation}\label{3.5.3}
\mbox{Tr}~e^{-tL_E}\sim {1 \over (4\pi t)^{D/2}}
\sum_{n=0}^\infty (A_n+A_{\beta,n})t^n.
\end{equation}
Here $A_n$ are standard coefficients
defined on the regular domain of ${\cal M}^E_\beta$.
Additions $A_{\beta,n}$ appear due to
conical singularities and accumulate
information about local geometry
near $\Sigma$. $A_{\beta,n}$ 
are local functionals
on $\Sigma$ expressed in terms of
powers of the Riemann tensor and
its derivatives projected with the help of $P^{\mu\nu}$. 
The quantities $\beta A_{\beta,n}$ are polynomial
in $\beta^{-1}$.
For the operator $L_E= -\nabla^2+\xi \breve{R}$ 
one has
\begin{equation}\label{3.5.5}
A_{\beta,0}=1~~,~~A_{\beta,1}=
{\pi \over 3\gamma}(\gamma^2-1)
\breve{\cal A},
\end{equation}
\begin{equation}\label{3.5.7}
A_{\beta,2}=
{\pi \over 3\gamma}\int_{\Sigma}
\left[{1 \over 60}(\gamma^4-1)\breve{\cal P}+(\gamma^2-1)\left(\frac 
16-\xi\right) \breve{R} \right],  
\end{equation} 
where $\gamma=2\pi/(\kappa\beta)$, $\breve{\cal A}$
is the surface area of Euclidean horizon $\Sigma$ 
and $\breve{\cal P}$ is the invariant on $\Sigma$ defined by 
(\ref{n6.2}).  Laplacians on $C_\alpha$ and 
the coefficient $A_{\beta,1}$ were discussed 
by Cheeger \cite{Cheeger:83}, (see also 
\cite{CKV:94},\cite{Fursaev:94a}).  The heat kernel expansion on spaces 
with fixed-point sets was analyzed by Donnely \cite{Donnelly:76}.  Heat 
kernel expansion (\ref{3.5.3}) on ${\cal M}^E_\beta$ has been studied 
in Refs. \cite{Fursaev:94b}, \cite{Dowker:94a}, \cite{Dowker:94b} where 
the explicit 
form of (\ref{3.5.7}) was found.  A summary of results for 
asymptotics for higher spin fields is presented in 
Ref. \cite{FF:98}.

\subsection{Infrared/ultraviolet duality}
\bigskip

Now we define the one-loop effective action $W^E(\beta)$ 
on ${\cal M}^E_\beta$ by using
dimensional regularization, i.e.
\begin{equation}\label{n6.3}
W^E(\beta)=\frac 12 \ln \det (L_E+m^2)
=-\int_0^{\infty}{dt \over t}
e^{-tm^2}
\mbox{Tr}~e^{-tL_E}.
\end{equation}
For $D\neq 4$ the singular part of (\ref{n6.3})
can be found from (\ref{3.5.3})
\begin{equation}\label{3.5.8}
W^E_{\mbox{div}}(\beta)=
-{1 \over 2}
\int_{0}^{\infty}{dt \over t}
{e^{-tm^2} \over (4\pi t)^{D/2}}
\left(B_0+t B_1+t^2 B_2\right),
\end{equation}
where $B_k=A_k+A_{\beta,k}$.
On a regular Euclidean manifold the ultraviolet
divergences are determined only
by $A_0$, $A_1$, $A_2$ which are proportional 
to the period $\beta$. 
So they appear only as divergencies
$E_{0,\mbox{div}}$ of the vacuum part $E_0$
of $F^E(\beta)$, see (\ref{main}).
In case of conical singularities
the ultraviolet divergences are polynomials in 
$\beta^{-1}$ because of 
$A_{\beta,k}$ and they also appear in the thermal part
of $F^E(\beta)$.
The divergent part $F^E_{\mbox{div}}(\beta,D)$ can
be found from 
(\ref{3.5.5}), (\ref{3.5.7}) and (\ref{3.5.8}).
Note that $F^E_{\mbox{div}}(\beta,D)$
is a local functional determined by curvatures and
geometrical invariants ($\breve{\cal A}$, 
$\breve{R}$, $\breve{\cal P}$)
at the Euclidean horizon.
For this local functional the Wick rotation
described in section 3.1, Eqs. (\ref{n.i7}),
has a well-defined meaning and it can be used 
to go in $F^E_{\mbox{div}}(\beta,D)$ 
from Euclidean to Lorenztian theory\footnote{This procedure is
analyzed in some detail in \cite{MaSo}
for a Kerr--Newman black hole.}.
As one can further show, after the 
Wick rotation  the thermal part of
$F^E_{\mbox{div}}(\beta,D)$ is identical to the 
divergent part of the canonical energy (\ref{n6.8a}).  
This coincidence  
also takes place  in the Pauli--Villars regularization, so one can 
write \begin{equation}\label{3.5.8a} 
F^E_{\mbox{div}}(\beta,\delta)-E_{0,\mbox{div}}(\delta)=
F^C_{\mbox{div}}(\beta,\delta),
\end{equation}
where $\delta$ is a regularization parameter 
($\delta=D-4$ for dimensional regularization
and $\delta=\mu^{-1}$ for PV regularization).

Relation (\ref{3.5.8a}) is remarkable. It demonstrates
a duality between large-volume infrared divergences 
which appear in $F^C(\beta)$ and ultraviolet-type divergences
which appear in $F^E(\beta)$ as a result of curvature
singularities on the Euclidean horizon. This duality is 
manifest in the Pauli--Villars regularization where the
inverse scale of Pauli--Villars  masses $\mu^{-1}$ plays 
the  same role as the cutoff parameter of integrals near the horizon.
 
\bigskip

We conclude our discussion with very brief comments
on the problem of statistical explanation of
the Bekenstein--Hawking entropy $S^{BH}$ of black holes.
Note  that the entropy 
of a field near the black hole horizon can be computed 
by the standard formula 
$S=\beta^2\partial_\beta F^C$ at the Hawking temperature
$\beta^{-1}=\kappa/(2\pi)$. By using (\ref{n6.12a}) 
one can show that to leading order
$S\sim {\cal A}/\delta^2$ where $\cal A$ is the surface
area of the black hole horizon and $\delta$ is the cutoff.  
The infrared/ultraviolet duality enables one to identify
$\delta^{-1}$ with a high-energy cutoff and,  in particular,
with the Planck mass.
This fact is interesting because it can be used 
as one of possible explanations 
of the Bekenstein--Hawking entropy $S^{BH}={\cal A}/(4G)$. 
Indeed, suppose that 
Einstein gravity is entirely induced by quantum effects of some 
fundamental underlying theory. For simplicity one can
assume that the degrees of freedom (constituents) of this theory are
some field variables. Then the induced Newton constant
is $G\sim \delta^2$ where $\delta^{-1}$ and 
$S^{BH}\sim {\cal A}/\delta^2$. Because of the duality property
the same cutoff $\delta$ which determines the induced Newton constant
appears also in the entropy of constituents $S$ and 
automatically $S$ has the same order of magnitude as $S^{BH}$. 
A similar mechanism may be realized in 
open string theory \cite{HMS}.
Further discussion of the black hole entropy problem in
induced gravity and concrete realizations of this idea
can be found in Ref. \cite{ind}.

\section{Resume}
\bigskip

The aim of these notes was to overview   
a number of new results and methods in finite-temperature
field theory in classical external stationary backgrounds.
We have also discussed their physical applications,
including quantum effects near rotating black holes. 
The physics here is reduced to a class of spectral problems which
depend on the spectral parameter in a non-linear way.
This also represents an interest form mathematical point of view 
because some results of the
spectral theory such as heat kernel asymptotics can be extended to this
class of  problems. One of our motivations was to find a connection
between canonical and Euclidean methods in statistical mechanics
in situations more non-trivial than static external fields.
It is interesting to note that the Euclidean gravity yields an
alternative derivation
of the spectral asymptotics. This enables one to avoid
a complicated analysis of inner products and measures
which has to be done in the Lorenztian theory. 
This fact hints that our results might be extended to a larger 
class of spectral problems.

\bigskip
\vspace{12pt} {\bf Acknowledgements}:\\

The author is grateful to I. Avramidi, G. Cognola, P. Gilkey,
A. Mostfazadeh, R. Seeley, and S. Zerbini
for interesting remarks regarding the spectral problems
discussed here. Many thanks go to G. Esposito for the suggestion
to write these notes and for proof reading 
of the manuscript.
This work was supported by the RFBR grant N 01-02-16791.

\end{document}